\documentclass[11pt,a4paper,english]{article}
\usepackage[T1]{fontenc}
\usepackage[latin9]{inputenc}
\usepackage{geometry}
\usepackage{amssymb}
\usepackage{graphicx}
\usepackage{epsfig}
\usepackage{subfigure}
\usepackage{babel}
\usepackage{amsmath}
\usepackage{color}
\usepackage{multicol,bbm}

\usepackage{jheppub}

\def\cR{{\mathcal R}}
\def\cB{{\mathcal B}}

\begin{document}

\title{Spectral curve for open strings attached to the $Y=0$ brane}

\author[a]{Zolt\'an Bajnok, } 
\author[a]{Minkyoo Kim, }  
\author[b]{L\'aszl\'o Palla, }  
 \affiliation[a]{
MTA Lend\" ulet Holographic QFT Group, Wigner Research Centre,
H-1525 Budapest 114, P.O.B. 49, Hungary
} 
\affiliation[b]{
Institute for Theoretical Physics, Roland E\"otv\"os University,
1117 Budapest, P\'azm\'any s. 1/A Hungary}


{\hfill ITP-Budapest Report 664}

\emailAdd{bajnok.zoltan@wigner.mta.hu}
\emailAdd{minkyoo.kim@wigner.mta.hu}
\emailAdd{palla@ludens.elte.hu}

\abstract{The concept of spectral curve is generalized to open strings in
    AdS/CFT with integrability preserving boundary conditions. 
    Our definition is based on the logarithms of the eigenvalues of the open monodromy
    matrix and makes possible to determine all
    the analytic, symmetry and asymptotic properties of the quasimomenta. 
    We work out the details of the whole construction for the $Y=0$ brane boundary condition.
    The quasimomenta of open circular strings are explicitly calculated. 
    We use the asymptotic solutions of the $Y$-system and the boundary Bethe Ansatz equations
    to recover the spectral curve in the strong coupling scaling limit. 
    Using the curve the quasiclassical fluctuations of some open string solutions are also studied.}

\maketitle

\setcounter{footnote}{0}

\section{Introduction}

One of the greatest progresses in contemporary theoretical 
physics is the AdS/CFT correspondence \cite{Maldacena:1997re, Gubser:1998bc, Witten:1998qj}. 
In the most analyzed 
version it relates the spectrum of IIB strings in the 
$AdS_5 \times S^5$ background to the scaling dimensions of 
single trace operators of the maximally supersymmetric
four dimensional gauge theory. The integrability, which 
shows up in the 't Hooft limit, allows a complete characterization 
and exact determination of the full spectrum \cite{Beisert:2010jr}. This characterization 
is different  in the weak and  strong coupling
regimes. 

In the strong coupling or (semi)classical domain 
integrability manifests itself by the existence of a spectral
parameter dependent flat connection \cite{Bena:2003wd}. The corresponding parallel
transporter can be evaluated on a non-trivial closed loop, 
which defines the monodromy matrix, whose 
trace is time independent and generates an infinite family of 
conserved charges. Even more, the logarithm of the eigenvalues of 
the monodromy matrix (quasi momenta) form an eight-sheeted 
Riemann surface: the spectral curve \cite{Beisert:2005bm}. The spectral curve provides
a very elegant description of the finite energy classical 
configurations. By requiring the right analytical properties
for the quasi momenta allows to find the classical curve for
each solution and to determine its energy without explicitly
constructing the solution itself \cite{Kazakov:2004qf}. Moreover, the curve can 
be used to characterize the small fluctuations around the classical solutions and by this way 
to describe their semiclassical corrections \cite{Gromov:2007aq}. 

The quantum spectrum of particles in a large volume can be 
described by specifying their dispersion relations and  
momentum quantization conditions, called the asymptotic 
Bethe Ansatz equations \cite{Beisert:2005fw}. These equations are valid for any 
coupling provided the volume is large. The obtained spectrum 
can be compared at weak coupling to the spectrum of the dilation 
operator of  gauge theory, while at strong coupling to the energies of the 
classical string solutions. 
The spectral curve can be also recovered in the 
strong coupling limit as Bethe roots condense and form the 
expected cuts \cite{pedro}. 

The complete quantum description of the spectrum valid for any
coupling and volume is given in terms of the Y-system \cite{Gromov:2009tv, Bombardelli:2009ns, Arutyunov:2009ur}.
The large volume solution of this Y-system is
related in a simple way to the asymptotic Bethe Ansatz 
equations and the spectral curve can be recovered in the strong 
coupling limit \cite{Gromov:2009tq, Gromov:2010vb}.  

The AdS/CFT correspondence which relates the scaling dimensions of 
single trace operators to the energies of closed string states  
relates also the scaling dimensions of 
determinant type operators and the energies of open string states. 
Open strings end on D-branes and a careful choice of the brane can 
ensure integrability in the 't Hooft limit \cite{Hofman:2007xp, Ahn:2008df}. 

The classical integrability of  open strings 
can be shown by constructing the analogue 
of the monodromy matrix. In the open case the parallel transporter
can be used to move from one boundary to the other. At the boundary 
gluing automorphism has to be introduced, such that when combined 
with the transporter a so called double row monodromy matrix is 
obtained \cite{Mann:2006rh, Dekel:2011ja}. Its trace is time independent and generates an infinite
family of conserved charges \cite{Dekel:2011ja}. 

The goal of the present paper is to generalize the spectral 
curve construction from the closed case to the open one and use it
to characterize the open string spectrum. We define the spectral curve
via the logarithm of the eigenvalues of the double row monodromy 
matrix. This definition determines the analytical properties of 
the spectral curve (including the cut structure, the poles with 
prescribed residues and its infinite asymptotics). 

We then analyze the curve form different point of views. For simplicity 
we restrict the investigations for the \(Y=0\) brane boundary conditions as 
in this case both the asymptotical Bethe Ansatz and the Y-system 
solutions are available \cite{Galleas:2009ye, Bajnok:2012xc}. 
We obtain and characterize the curve as 
the semi-classical limit of these descriptions. We also construct
explicitly the spectral curve for the BMN state and for 
circular open strings. Finally, we show how the spectrum of small 
fluctuations can be determined. 

The paper is organized as follows: We start in the next section 
by the Lagrangian definition of the model. We follow the notation
of \cite{Dekel:2011ja}, where the monodromy matrix was constructed. We then 
introduce the quasimomenta and list its properties such as 
symmetries, asymptotics and singularity structure. In Section 
3 we provide the explicit BMN and circular strings solutions and 
calculate the quasi momenta from first principles.  
Section 4 shows how one can derive the spectral curve from the boundary Y-system, 
while section 5 contains the analogous 
derivation starting from the asymptotic BA equations. 
We analyze the small fluctuations in the language of the spectral curve in section 6.
Finally we  conclude in section 7. The details of the calculations are relegated to Appendices. 

\section{The \lq\lq open'' monodromy matrix and the quasimomenta} 

In this section we define the spectral curve for open strings from
the logarithmic derivative of the eigenvalues of the boundary 
monodromy matrix. 

\subsection{Monodromy matrix}
The boundary monodromy matrix is the analogue of the periodic 
monodromy matrix and generates an infinite family of conserved 
charges.

\subsubsection{Flat connections and integrability}

Classically the open superstring on $AdS_5\times S^5$ 
coupled to the $Y=0$ brane is described with
the help of the Green-Schwartz sigma model (GS$\sigma $M) taking
 values in $\mathrm{su}(2,2\vert4)$
\footnote{Our coupling constant is defined as 
$g=\frac{\sqrt{\lambda}}{4\pi}$ with $\lambda$ being the 't Hooft coupling.}:
\begin{equation}
S=-g\int d\tau d\sigma [\gamma^{\alpha\beta}\mathrm{str}(A^{(2)}_\alpha
  A^{(2)}_\beta )+k \epsilon^{\alpha\beta}\mathrm{str}(A^{(1)}_\alpha
  A^{(3)}_\beta )],\quad k =\pm 1
\end{equation}
where $A^{(i)}$ denote the various $\mathbb{Z}_4$ components of 
the Maurer-Cartan one form $A$:
\begin{equation}
A=-g^{-1}dg=\sum\limits_{i=0}^3A^{(i)},\qquad g\in SU(2,2\vert4)
\end{equation} 
In contrast to the closed string (periodic) case this sigma model has some non trivial 
but consistent and integrability preserving  boundaries \cite{Dekel:2011ja}.   
 The integrability of the model is guaranteed by the existence of the
\lq\lq moving frame'' flat connection, $L_\alpha$ ($\alpha=\tau,\ \sigma$): 
\begin{equation}
L_\alpha=l_0A_\alpha^{(0)}+l_1A_\alpha^{(2)}+l_2
\gamma_{\alpha\beta}\epsilon^{\beta\rho}A_\rho^{(2)}
+l_3A_\alpha^{(1)}+l_4A_\alpha^{(3)}\,,\qquad \gamma_{\alpha\beta}=\rm{diag}(-1,1)
\label{nagyL}
\end{equation}
where the $l_i$ parameters are obtained from requiring the
equations of motions for $A^{(i)}$ to coincide with the conditions of 
vanishing curvature \cite{Bena:2003wd, Arutyunov:2009ga}. 
They can be written in terms of a complex variable
$\zeta$ as:
\begin{equation}
l_0=1,\qquad l_1=\frac{1}{2}(\zeta^2+\zeta^{-2}),\qquad
l_2=-\frac{1}{2 k}(\zeta^2-\zeta^{-2}),\qquad l_3=\zeta,\qquad
l_4=\zeta^{-1}.
\label{liparams}
\end{equation}  
The integrability preserving boundary conditions can be nicely formulated in terms of the
\lq \lq fixed'' frame connection $l_\alpha$, which is the gauge transform of $L_\alpha$:
\begin{equation}
l_\alpha=gL_\alpha g^{-1}+\partial_\alpha g\,g^{-1}
\end{equation} 
and is correspondingly flat itself:
\begin{equation}
dl(\zeta)-l(\zeta)\wedge l(\zeta)=0~.
\end{equation}
By introducing  $\sum\limits_{i=0}^3gA^{(i)}g^{-1}=\sum\limits_{i=0}^3a^{(i)}$ the 
fixed frame $l_\alpha$ can be written as
\begin{equation}
l_\alpha=(l_1-1)a_\alpha^{(2)}+l_2\gamma_{\alpha\beta}
\epsilon^{\beta\rho}a_\rho^{(2)}+(l_3-1)a_\alpha^{(1)}+(l_4-1)a_\alpha^{(3)}\,.
\label{lexplicit}
\end{equation}
The integrability preserving boundary conditions are given by appropriate gluing conditions 
on this fixed frame flat connection  ($l(\zeta)\equiv l_\tau(\zeta)d\tau+l_\sigma(\zeta)d\sigma$) as 
\begin{equation}
l(\zeta)=\Omega(\bar{l}(\zeta^{-1})) \quad {\rm at} \quad \sigma =0,\ \pi,
\label{bc}
\end{equation}
where $\Omega$ is an involutive metric preserving automorphism  and ${\bar l}(\zeta)\equiv l_\tau(\zeta)d\tau-l_\sigma(\zeta)d\sigma$. These
conditions guarantee, that the boundary terms arising in the variation of the
action as a result of the open ends just cancel. 

\subsubsection{Boundary monodromy matrix}

Similarly to the closed string case the generator of the conserved quantities is described 
through the transport matrix
\begin{equation}
T(\sigma_2,\sigma_1,\zeta)=P\exp\bigl(\int_{\sigma_1}^{\sigma_2}d\sigma
l_\sigma(\sigma, \zeta)\bigr)
\end{equation}
In the periodic case, when $l(0,\zeta)=l(2\pi,\zeta)$, the generator is given by 
${\rm Str}( T_\gamma(\zeta))$, where $T_\gamma$ (the \lq\lq closed'' monodromy matrix) is
the transport matrix around the cylindrical world-sheet
$T_\gamma(\zeta)=T(2\pi,0,\zeta)$.  In the presence of boundaries the authors of
\cite{Dekel:2011ja}  define the open monodromy matrix as
\begin{equation}
T(\zeta)=U_0T^{-1}(\pi,0,\zeta^{-1})U_\pi T(\pi,0,\zeta)
\label{Tdef}
\end{equation}
where $U_{0,\pi}$ are constant matrices (\lq\lq classical reflection
matrices'') with $U_{0,\pi}^2 =\pm 1$,  and show that the supertrace
of the monodromy matrix is time-independent, 
\begin{equation}
\partial_\tau {\rm Str}(T(\zeta))=0\quad \longleftrightarrow 
\quad U_i l_\tau(i,\zeta)U^{-1}_{i}=l_\tau(i,\zeta^{-1})\quad ,\quad i=0,\pi 
\end{equation}
provided the automorphism $\Omega_U (h)=U\, h\, U^{-1}$ has the appropriate
properties eq.(\ref{bc}))\footnote{For the giant graviton/$Y=0$ brane the $U_{0,\pi}$ matrices are
  given explicitly in Appendix \ref{sec:Uexplicit}.}. Thus, in the case of integrable boundaries, we can think of 
${\rm Str}(T(\zeta))$ as the classical version of the {\sl double row
  transfer matrix}, which generate the conserved charges. 

\subsubsection{Monodromy matrix and conserved charges}
  
It is important to obtain the relation between the \lq\lq open'' monodromy
matrix $T(\zeta)$ and the conserved global charges $Q$. To see this
one expands $l_\sigma(\zeta )$ and $T(\zeta)$ around
$\zeta=1$. Writing $\zeta=1-w$ in (\ref{lexplicit}) 
it is straightforward to show that
$l_\sigma(1-w)=wJ^\tau /g +\dots$, where $J^\tau $ is the time-like component of
the conserved $\partial_\alpha J^\alpha =0$ global symmetry current 
\cite{Arutyunov:2009ga} thus 
$T(\pi,0,1-w)=1+wQ/g+\dots $ and 
\begin{equation}
T(\zeta)\vert_{\zeta=1-w}=U_0U_\pi+\frac{w}{g}(U_0QU_\pi+U_0U_\pi
Q)+\dots =U_0U_\pi(1+\frac{2w}{g}Q+\dots)\,.
\label{megmarado}\end{equation} 
Here, in the last equality, we exploited that $[Q,U]=0$ must hold for the
conserved charges to be consistent with the boundary condition (\ref{bc}).

\subsubsection{Symmetries of the monodromy matrix}

The symmetry equations for $T(\zeta)$ are obtained by combining the
transformation property of $L_\alpha(\zeta)$ under the $\mathbb{Z}_4$
automorphism \cite{Arutyunov:2009ga}
\begin{equation}
\mathcal{K}L_\alpha(\zeta )^{\rm{ST}}\mathcal{K}^{-1}=-L_\alpha(i\zeta )
\label{Ltransform}
\end{equation}
(where $\mathcal{K}$ is the $8\times 8$ matrix implementing the automorphism)
and the $[U_0,g(0)]=0$, $[U_\pi,g(\pi)]=0$ properties \cite{Dekel:2011ja} of the 
$U_{0,\pi}$ matrices. 

First we relate $T(\zeta)$ to the analogous open
monodromy matrix built with the aid of the $L_\alpha$ connection instead of
$l_\alpha$: denoting $\tilde{T}(\pi
,0,\zeta)=P\exp(\int\limits_0^\pi d\sigma L_\sigma(\zeta))$ we find
\begin{equation}
T(\zeta)=g(0)^{-1}\tilde{T}(\zeta)g(0),\quad
\tilde{T}(\zeta)=U_0\tilde{T}^{-1}(\pi,0,\zeta^{-1})U_\pi
\tilde{T}(\pi,0,\zeta )\,.
\label{Llrelation}
\end{equation}
Then, since according to \cite{Dekel:2011ja} $U_{0,\pi}$ also satisfy 
$\mathcal{K}^{-1}U_{0,\pi}\mathcal{K}=-U_{0,\pi}^{\rm{ST}}$, using also
(\ref{Ltransform}),  
one easily gets
\begin{equation}
\tilde{T}(i\zeta)=\mathcal{K}\Bigl(\tilde{T}^{-1}(\zeta)
\Bigr)^{\rm{ST}}\mathcal{K}^{-1}\,,
\end{equation}
It follows then that $T(\zeta)$ satisfies the symmetry equation
\begin{equation}
T(i\zeta)=\tilde{\mathcal{K}}(T^{-1}(\zeta))^{\rm{ST}}(\tilde{\mathcal{K}})^{-1},\qquad
\tilde{\mathcal{K}}=g(0)^{-1}\mathcal{K}(g(0)^{-1})^{\rm{ST}}
\label{sym1}
\end{equation}
while the definition and $U_{0,\pi}^2=\pm 1$ guarantee that
\begin{equation}
T(\zeta^{-1})=U_0 T^{-1}(\zeta) U_0^{-1}
\label{sym2}
\end{equation}
is also satisfied.

\subsection{The spectral curve of quasi-momenta}

In the following we define the spectral curve from the eigenvalues of the  \lq\lq open''  
monodromy matrix $T(\zeta)$. 

\subsubsection{Quasi-momenta}

The $4+4$ eigenvalues $(\lambda_1,\dots
\lambda_4\vert\mu_1,\dots \mu_4)$ of  $T(\zeta)$ can be expressed in terms of the so called 
quasi-momenta of 
$S^5$ and $AdS_5$ as
\begin{equation}
\lambda_i=e^{-i\tilde{p}_i(\zeta)} ,\qquad \mu_i=e^{-i\hat{p}_i(\zeta)},\qquad i=1,\dots 4
\label{qmomenta}
\end{equation}
Following from its definition $T(\zeta)$ depends analytically on
$\zeta$ (apart from the points 
$\zeta=0,\ \infty$), but this property is not necessarily inherited by the $\lambda_i$
$\mu_i$ eigenvalues. Just as in the closed string case \cite{Beisert:2005bm}
there are square root type singularities  
when two $\lambda$-s or two $\mu$-s coincide while at those points where
eigenvalues having opposite gradings coincide both of them have first order
poles. To obtain a single valued and analytic function on the entire complex
plane with the exception of these singularities we define $Y(\zeta )$ 
- in analogy to the
closed string case \cite{Beisert:2005bm} - as
\begin{equation}
m(\zeta)Y(\zeta)m^{-1}(\zeta)=-i\zeta\frac{d}{d\zeta}\log(m(\zeta)T(\zeta)m^{-1}(\zeta)) 
\end{equation}
where $m(\zeta)$ diagonalizes $T(\zeta)$. This definition makes it
possible to write\footnote{This form shows that $Y(\zeta)$ contains only pole type 
singularities since $M(\zeta)$ has first order poles  
at the square root type branch points of $m(\zeta)$.}
\begin{equation}
Y(\zeta)=
T^{-1}(\zeta)\Bigl(-i\zeta\frac{d}{d\zeta}T(\zeta)+[M(\zeta),T(\zeta)]\Bigr),\qquad
M(\zeta)=-i\zeta m^{-1}\frac{d}{d\zeta}m\,.
\label{ytm}
\end{equation}  
 The eigenvalues of
$Y(\zeta)$ (that are the logarithmic derivatives of $\lambda_i$ and $\mu_i$)
are determined by the zeroes and poles of its characteristic function
\begin{equation}
F(\tilde{y}(\zeta),\zeta)=0,\qquad F(\hat{y}(\zeta),\zeta)=\infty,\qquad
F(y,\zeta)=\frac{\tilde{P}(\zeta)}{\hat{P}(\zeta)}{\rm{sdet}}\Bigl(y-Y(\zeta)\Bigr).
\end{equation}
(The polynomial prefactors are introduced to absorb the poles coming from
$M(\zeta)$ without changing the curve see \cite{Beisert:2005bm}). The symmetry
equations of the \lq\lq open'' monodromy matrix (\ref{sym1}, \ref{sym2}) can be
converted into symmetry equations of $Y(\zeta)$ and $M(\zeta)$
\begin{eqnarray}
&& Y(i\zeta)=-\tilde{\mathcal{K}}Y(\zeta)^{\rm{ST}}\tilde{\mathcal{K}}^{-1},\quad \,\
M(i\zeta)=-\tilde{\mathcal{K}}M(\zeta)^{\rm{ST}}\tilde{\mathcal{K}}^{-1}, \label{yszimmetria}\\
&& Y(\zeta^{-1})= \mathcal{N} Y(\zeta) \mathcal{N}^{-1},\qquad
M(\zeta^{-1})=-U_0 M(\zeta) U_0^{-1},
\label{yszimmetria1}
\end{eqnarray}
(with $\mathcal{N}=U_0T(\zeta)$) and these equations imply that
\begin{equation}
F(y,i\zeta)=F(-y,\zeta),\qquad {\rm and}\qquad F(y,\zeta^{-1})=F(y,\zeta).
\label{fszimmetria}\end{equation} 
Therefore $F(y,\zeta)$ may depend analytically only on $y^2$,
$y(\zeta^2+\zeta^{-2})$ and $\zeta^4+\zeta^{-4}$, and $y$ must be a function
of $\zeta^2+\zeta^{-2}$. We introduce the variable 
\begin{equation}
x=\frac{1+\zeta^2}{1-\zeta^2}
\end{equation}
which is identical to the spectral parameter used in the closed string case 
\cite{Beisert:2005bm} - and from now on we may think of (the eigenvalues of) the
open monodromy matrix as being a function of $x$:  
$T(x)$. (Note that $\zeta\rightarrow 1/\zeta$ changes $x$ as
$x\rightarrow -x$, while the $\zeta\rightarrow i\zeta$ map induces 
$x\rightarrow 1/x$).  

\subsubsection{Symmetries and analytical structure of the quasi-momenta}

The symmetry equations (\ref{yszimmetria}, \ref{yszimmetria1}, \ref{fszimmetria}) impose some
restrictions on the quasimomenta $(\tilde{p}_i,\hat{p}_i)$. Since the 
$x\rightarrow 1/x$ 
 ($\zeta\rightarrow i\zeta $) inversion symmetry equations
for $Y(\zeta )/F(y,\zeta )$ are the same as for the closed string case the
restrictions they impose are also the same (with a minor difference):
\begin{eqnarray}
&& \tilde{p}_{1,2}(x)=-\tilde{p}_{2,1}(1/x),\quad 
\tilde{p}_{3,4}(x)=-\tilde{p}_{4,3}(1/x),\quad \\ \nonumber
&& \hat{p}_{1,2}(x)=-\hat{p}_{2,1}(1/x),\quad 
\hat{p}_{3,4}(x)=-\hat{p}_{4,3}(1/x),
\end{eqnarray}
where the absence of winding in the $S^5$ component $\tilde{p}_i$ is the
difference to the closed string case. On the other hand the $x\rightarrow -x$  
 ($\zeta\rightarrow 1/\zeta $) reflection symmetry equations in 
(\ref{yszimmetria1}, \ref{fszimmetria}) require
\begin{equation}
 \tilde{p}_{i}(-x)=-\tilde{p}_{i}(x),\quad
\hat{p}_{i}(-x)=-\hat{p}_{i}(x),\quad i=1,\dots ,4.
\end{equation}These extra properties are the consequence of the boundaries and are not present for generic periodic states. Indeed, the quasimomenta ${\hat p}_{i}$ are related to the even eigenvalues $(y_{i}(-x)=y_{i}(x))$ of $Y(x)$ by $(x^2 -1) \frac{dp_i}{dx} = y_{i} (x)$ since $ \zeta \frac{d}{d\zeta} = (x^2-1) \frac{d}{dx}$ and we choose the integration constant to guarantee (\ref{fszimmetria}). Alternatively, the reflection symmetry of the quasimomenta can be directly obtained from (\ref{sym2}). 

The $l_\alpha $ (or $L_\alpha $) connection has singularities at $x=\pm 1$
($\zeta =0$ resp. $\zeta =\infty$) and they imply simple poles for the
quasimomenta. $l_\alpha $ is supertraceless since $l_\alpha \in
\mathrm{psu}(2,2\vert4)$ while the Virasoro constraint - which is not modified
by the presence of the boundary - forces its square also to be
supertraceless. Combining these with the inversion and reflection symmetries
synchronizes the various residua as:
\begin{equation}
\{\tilde{p}_1,\tilde{p}_2,\tilde{p}_3,\tilde{p}_4\vert 
\hat{p}_1,\hat{p}_2,\hat{p}_3,\hat{p}_4 \}\sim\frac{x}{x^2 -1}\{ \alpha
,\alpha ,\beta ,\beta \vert \alpha ,\alpha ,\beta ,\beta \}
\label{polszink}\end{equation}

Finally we mention that the asymptotic behaviour in (\ref{megmarado}) can be
converted into the $x\rightarrow\infty $ behaviour of the quasimomenta
\begin{equation}
\mathrm{diag}(\tilde{p}_1,\tilde{p}_2,\tilde{p}_3,\tilde{p}_4\vert 
\hat{p}_1,\hat{p}_2,\hat{p}_3,\hat{p}_4 )\sim \frac{2}{gx}iQ_{\mathrm{diag}}
\end{equation}
where $Q_{\mathrm{diag}}$ is the sum of Cartan generators with eigenvalues
characterising the solution. Note that they automatically commute with the diagonal
$U$, eq.(\ref{Uexplicit}), of the $Y=0$ brane.  

The $\{\tilde{p}_i(x)\vert\hat{p}_i(x)\}$ quasimomenta form an eight sheeted
Riemann surface very similar to the closed string case, where the
$\tilde{p}_i(x)$ and $\hat{p}_i(x)$ sheet functions are analytic almost
everywhere. Apart from the single poles at $x=\pm 1$ where their residua are
synchronized  as in (\ref{polszink}) they may have branch cuts with square
root type end points connecting either two $\tilde{p}_i(x)$ or two 
$\hat{p}_i(x)$ sheets (corresponding to bosonic degrees of freedom) or they
may have single poles existing simultaneously on a $\tilde{p}_i(x)$ and a 
$\hat{p}_j(x)$ sheet (corresponding to fermionic degrees of freedom). The
important point is that these cuts and poles must respect the inversion and
reflection symmetries: the (non invariant) generic ones come in fourfold
multiplets to provide the representation of the symmetry.

\section{Explicit quasimomenta for circular open strings \label{sec3}}  

The metric on $AdS_5\times S^5$ is given by 
\begin{eqnarray}
ds^2 &=& -\cosh^2\rho dt^2+d\rho^2+\sinh^2\rho(d\alpha^2+\sin^2\alpha
d\Phi^2+\cos^2\alpha d\phi^2)\nonumber\\
     &+&d\gamma^2+\cos^2\gamma d\phi_1^2+\sin^2\gamma (d\psi^2+\cos^2\psi
d\phi_2^2+\sin^2\psi d\phi_3^2)
\end{eqnarray}
while the global $X$, $Z$ and $Y$ coordinates of the $S^5$ are 
\begin{equation}
X=\cos\gamma e^{i\phi_1}\qquad Z=\sin\gamma\cos\psi e^{i\phi_2}\qquad 
Y=\sin\gamma\sin\psi e^{i\phi_3} 
\label{globalS5}
\end{equation}
The giant graviton corresponding to the $Z=0$ brane ($\psi\equiv \pi/2$) 
is the $S^3$ described by
\begin{equation}
d\gamma^2+\cos^2\gamma d\phi_1^2+\sin^2\gamma d\phi_3^2
\end{equation}
 while for the $Y=0$ brane 
($\psi\equiv 0$ or $\psi\equiv\pi$) it is the $S^3$ given by 
\begin{equation}
d\gamma^2+\cos^2\gamma d\phi_1^2+\sin^2\gamma d\phi_2^2\,. 
\end{equation}
These two $S^3$-s are of course obtained from each other by a rotation,
however the open strings ending on them have different properties because the
two $S^3$-s are aligned in a different way with respect to the ground
state. In the following we will construct the explicit quasimomenta for $Y=0$ brane set-up. 

We first recall from \cite{Stefanski:2003qr}, that a simple rotating spinning closed string solution
when cut into ``half'' satisfies the boundary conditions 
\begin{equation}
Y=0,\qquad \partial_\sigma X=\partial_\sigma Z=0
\end{equation}
appropriate for the $Y=0$ brane, thus the half can be used as a rotating
spinning open string solution. In the simplest case this solution just
describes the open BMN string. We investigate the open monodromy matrix
$T(x)$ for a class of solutions and obtain its eigenvalues
explicitly not only for the BMN string but also for a subset of the rotating
spinning strings.

This class of solutions is given by $\rho\equiv 0$ and 
the following $X$, $Y$ and $Z$ 
\begin{equation}
\gamma\equiv\frac{\pi}{2}\quad\leftrightarrow\quad X\equiv 0,\qquad
Z=\cos(n\sigma)e^{iw\tau},\qquad Y=\sin(n\sigma)e^{iw\tau}\qquad t=\kappa\tau\,,
\end{equation}
where $\sigma$ is running in $(0,\pi)$ only, $n$ is an integer (we consider  
the case when it is an even integer), and the $\kappa$, $w$ constants are given
as  
\begin{equation}
w^2=n^2+\nu^2,\qquad \kappa^2=\nu^2+2n^2    
\end{equation}
in terms of $n$ and an arbitrary real constant $\nu$. (Note that the interior
of the open string is away from the $Y=0$ surface, only its endpoints move
there). 
The energy and angular
momenta of this open string solution are \footnote{Note that $L=J_Z+J_Y$ corresponds to the number of fields in the determinant type operators of $N=4$ SYM. For example, in the operator such as ${\cal O} \sim \sum_{k} \epsilon_{i_1 \ldots i_N} \epsilon^{j_1 \ldots j_N} Y_{j_1}^{i_1} \ldots Y_{j_{N-1}}^{i_{N-1}} (Z^k \chi Z^{J-k})_{j_N}^{i^N} $, the complex scalar $Z$ and impurity $\chi$ correspond to the open strings configuration and the part of $Y$ products is the maximal giant graviton. In this example, the total number of $Z$ and $\chi$ fields is $L=J+1$. } 
\begin{equation}
E=\frac{1}{2}\sqrt{\lambda}\kappa,\qquad
J_Z=J_Y=\frac{1}{4}\sqrt{\lambda}\sqrt{n^2+\nu^2}\,.
\end{equation}
The BMN string is obtained for $n=0$, in this case $E$ and $J_{Z,Y}$ become
proportional to each other. 

The first step to construct the open monodromy matrix for this solution is to
obtain the explicit form of the bosonic sectors current  (which is,
in fact the complete current since the solution has no fermionic component). 
We can do this in two ways: either we specialize the general expression in 
\cite{Dekel:2011ja} to the present case or we construct it from the coset space representative
appropriate for the solution:
$g_{\rm sol}=e^{-P_0\kappa\tau}e^{P_8w\tau}e^{J_{56}w\tau}e^{P_6n\sigma}$. See 
Appendix A for the explicit $P_i$ matrices. In either way
one obtains
\begin{equation}
A^{(2)}_\tau=P_0\kappa+P_5w\sin(n\sigma)+P_8w\cos(n\sigma),\qquad\quad 
A^{(2)}_\sigma=P_6n,\qquad\quad A^{(0)}_\sigma\equiv 0.
\end{equation}
Through eq.(\ref{nagyL}-\ref{liparams}) this leads to
\begin{equation}
L_\sigma=\frac{x^2+1}{x^2-1}P_6n-\frac{2x}{x^2-1}\Bigl(
P_0\kappa+P_5w\sin(n\sigma)+P_8w\cos(n\sigma)\Bigr)\,.
\end{equation}
The relation between $T(x)$ and the analogous expression built
with the aid of $L_\sigma$ instead of $l_\sigma$ is given in
eq.(\ref{Llrelation}), where $g(0)=g_{\rm sol}(0)=
e^{-P_0\kappa\tau}e^{P_8w\tau}e^{J_{56}w\tau}$. Since this is a similarity
transformation as long as we are interested in the eigenvalues of 
$T(x)$ we may consider that of $\tilde{T}(x)$
instead. 

\subsection{The BMN string} 

For the BMN string, when $n=0$, the situation is even simpler as 
\begin{equation}
l_\sigma=-\frac{2x}{x^2-1} \nu (P_0+P_8)
\end{equation}
is independent of $\sigma$ thus $T(\pi, 0,x)$ is readily obtained
\begin{equation}
T(\pi, 0,x)=\exp\Bigl(\Omega (P_0+P_8)\Bigr),\qquad
\Omega=-\frac{ 2\pi \nu x}{x^2-1}\,.
\end{equation}
This then leads through (\ref{Tdef}) and (\ref{Uexplicit}) to
\begin{equation}
T(\zeta )=(-)\begin{pmatrix} M & \\
 & {\rm diag}(e^{i\Omega},e^{i\Omega},e^{-i\Omega},e^{-i\Omega})
\end{pmatrix}\,,\quad 
M=\begin{pmatrix}
        \cos\Omega & 0 & \sin\Omega & 0 \\
         0 & \cos\Omega & 0 & -\sin\Omega \\
         -\sin\Omega & 0 & \cos\Omega & 0 \\
         0 & \sin\Omega & 0 & \cos\Omega
   \end{pmatrix}\,.
\label{TBMN}
\end{equation}
Note that this open monodromy matrix is non diagonal even for the BMN string,
since $M$ is non diagonal. Nevertheless $M$-s eigenvalues - two times
$e^{i\Omega}$ and two times $e^{-i\Omega}$ - coincide with those in the lower
right corner of $T(x )$.    
The quasimomenta for the BMN string as one reads off from (\ref{TBMN}) are
\footnote{Note that these quasimomenta are identical to the ones of the closed 
BMN string.} 
\begin{equation}
\hat{p}_{1,2}=-\hat{p}_{3,4}=\tilde{p}_{1,2}=-\tilde{p}_{3,4}=\frac{2\pi\nu
  x}{x^2 -1}\,,
\label{BMNmomenta}\end{equation}

\subsection{Quasimomenta for the solutions with $n=2N\neq 0$}

As mentioned above for these solutions we determine the eigenvalues of the open monodromy
matrix built from the $L_\sigma $ connection instead of $l_\sigma$ 
\begin{equation}
\tilde{T}(x)=U_0\tilde{T}^{-1}(\pi,0,-x)U_\pi
\tilde{T}(\pi,0,x ),\quad\qquad 
\tilde{T}(\pi
,0,x)=P\exp(\int\limits_0^\pi d\sigma L_\sigma(x)),
\end{equation}
since - according to eq.(\ref{Llrelation}) - they are the same as that of
$T(\zeta )$. 

We start with the matrix form of $L_\sigma$ 
\begin{equation}
L_\sigma= \frac{x^2 +1}{x^2 -1}P_6n-
\frac{2x}{x^2 -1}\Bigl(
P_0\kappa+P_5w\sin(n\sigma)+P_8w\cos(n\sigma)\Bigr)=\begin{pmatrix}H & 0\\
                                                                  0 &
                                                                  K\end{pmatrix},\end{equation}
where $H$ and $K$ are $4\times 4$ matrices. The $K$ matrix in the \lq\lq
$AdS_5 $ corner'' is diagonal
\begin{equation}
K=-\frac{2x}{x^2 -1}
P_0\kappa =-\frac{2x}{x^2 -1}\kappa\frac{i}{2}\mathrm{diag}\,(1,1,-1,-1),
\end{equation}
thus the $AdS_5$ eigenvalues of $T(x )$ are the doubly degenerate 
$e^{\pm
  i\frac{2\pi\kappa x}{x^2 -1}}$. This leads to the following $AdS_5$
quasimomenta
\begin{equation}
\hat{p}_{1,2}=-\hat{p}_{3,4}= \frac{2\pi\kappa x}{x^2 -1}=\frac{x}{x^2 -1}\frac{E}{g}.
\label{adsmom}\end{equation}

The matrix $H$ in the \lq\lq $S_5$ corner'' of $L_\sigma $  is 
\begin{equation}
H=\begin{pmatrix}0 & -\tilde{b}\\
                 \tilde{b} & 0\end{pmatrix},\quad\tilde{b}=
\tilde{\beta}\begin{pmatrix} xw(e^{in\sigma}+e^{-in\sigma}) &
  \hat{n}-xw(e^{in\sigma}-e^{-in\sigma})\\
\hat{n}+xw(e^{in\sigma}-e^{-in\sigma}) &
-xw(e^{in\sigma}+e^{-in\sigma})\end{pmatrix}
\label{S5corner}\end{equation}
where $\tilde{\beta}=\frac{1}{2(x^2-1)},\quad\hat{n}=n(x^2 +1) $. The problem with 
this matrix 
is that it depends on a non trivial way on $\sigma $,  which makes very 
complicated to compute its path ordered exponential 
$ \tilde{t}(\pi
,0,x)=P\exp(\int\limits_0^\pi d\sigma H(x))$. We overcome this problem by
recalling that this path ordered exponential is related to the solution of the
(vector) differential equation $\partial_\sigma\psi =H\psi$ by 
$\psi(\sigma)=\tilde{t}(\sigma ,0,x)\psi (0)$. We solve this linear 
problem in Appendix \ref{sec:Teigenv} and obtain the following $S^5$ quasimomenta:
\begin{equation}
\tilde{p}_1=\frac{2\pi x}{x^2
  -1}\sqrt{\frac{n^2}{x^2}+w^2}=-\tilde{p}_4,\qquad   \tilde{p}_2=\frac{2\pi
  x}{x^2 -1}\sqrt{n^2x^2+w^2}=-\tilde{p}_3.
\label{s5mom}\end{equation}
(Note that these quasimomenta are identical to the \lq\lq one cut'' solutions
in \cite{Kazakov:2004qf, Gromov:2009zz} ). The set of
quasimomenta given in eq.(\ref{adsmom}, \ref{s5mom}) satisfies the requirements
following from the inversion and reflection symmetries as well as the residuum
synchronization condition in a non trivial way.

\section{Quasimomenta from the $Y$ system}

For the periodic, closed string case the solutions of the AdS/CFT $Y$ and $T$
systems are obtained in the strong coupling scaling limit in
\cite{Gromov:2009tq, Gromov:2010vb}. These limiting solutions can be
compared to the result of semiclassical quantization based on the spectral
curve, in particular, the classical quasimomenta could be described in terms of
some conserved quantities and certain densities of Bethe roots (resolvent
densities).   In a recent paper \cite{Bajnok:2012xc} a conjecture is made 
that the $Y=0$ brane is described by
the same $Y$ and $T$ systems as the closed (periodic) case, only the
asymptotic solutions and the analytic properties of the $Y$ and $T$
functions  are different. Therefore, in this section, 
 we repeat the procedure of
\cite{Gromov:2009tq} and \cite{Gromov:2010vb} for the $Y=0$ brane, i.e.
we consider the strong
coupling scaling limit of the asymptotic ($L\rightarrow\infty$) $Y$ and $T$
functions described in \cite{Bajnok:2012xc} and obtain the quasimomenta 
from the limiting
solution.     

\subsection{Classical T-system for the $Y=0$ brane}   
 
First, following \cite{Gromov:2010kf},     
we consider the monodromy matrix $T(x)$ in appropriate unitary highest
weight irreps $\Lambda$ of $SU(2,2\vert 4)$ to describe 
the classical T-system, and denote their supertrace as  
$D_\Lambda={\rm Str}_\Lambda T(x)$. The irreps having rectangular
Young-tableaux 
$h_i=s+2$, $i=1,\dots a$ (denoted as $[a,s]$) play a distinguished role, as
they form a closed set under tensor multiplication:
$[a,s]\otimes [a,s]=[a+1,s]\otimes [a-1,s]\oplus [a,s-1]\otimes
[a,s+1]$. Evaluating this equation for the representatives of the monodromy
matrix and taking the supertrace we find
\begin{equation}
D_{a,s}D_{a,s}=D_{a-1,s}D_{a+1,s}+D_{a,s-1}D_{a,s+1}
\label{cHirota}
\end{equation}
This classical equation is the strong coupling $g\rightarrow\infty$ limit of 
the quantum Hirota equation: 
\begin{equation}
\mathbb{D}_{a,s}^+\mathbb{D}_{a,s}^-=\mathbb{D}_{a-1,s}\mathbb{D}_{a+1,s}
+\mathbb{D}_{a,s-1}\mathbb{D}_{a,s+1}
\label{qHirota}
\end{equation}
where here and from now on $f^{[n]} (u)=f(u+n\frac{i}{2})$ ;  $f^\pm(u)\equiv
f^{[\pm 1]}(u)$. 
In this limit the left hand side of the latter 
equation contains no shift in the parameter $u=g(x+1/x)$ (since $u\sim g$). 
Note that in the case of closed strings $T_{a,s}={\rm
  Str}_{[a,s]} T_\gamma(x)$ satisfy the same equations (\ref{cHirota}). This
gives further 
support to the conjecture made in \cite{Bajnok:2012xc}. 

The general solution
of equations (\ref{cHirota}) 
(with appropriate \lq\lq T hook'' boundary conditions) is
given in \cite{Gromov:2010vb}.  
  
\subsection{Asymptotic $Y$ and $T$ functions in the
  scaling limit}

We collect here the asymptotic large $L\rightarrow\infty$  solutions of the 
$Y$ and $T$ (quantum Hirota eq.(\ref{qHirota})) systems as given for the $Y=0$ brane in
\cite{Bajnok:2012xc}\footnote{These solutions are also called asymptotic $Y$
  and $T$ functions.}.
  
For each state in the theory there is a collection of $Y$ functions satisfying 
the $Y$ system relation
\begin{equation}
Y_{a,s}^+Y_{a,s}^-=\frac{(1+Y_{a,s-1})(1+Y_{a,s+1})}{(1+1/Y_{a-1,s})(1+1/Y_{a+1,s})}
\end{equation}
where $a$ and $s$ are integers.  Non-trivial $Y$ functions live on the 
 \lq\lq T-hook'' in which either $a=1$ or $s \in (-1,0,1)$ and $a$ positive. 
There are also two  \lq\lq exceptional'' points $a=2,s=\pm 2$. 
  
This $Y$ system can be solved in terms of the $T$ system, whose elements in 
the boundary problem are denoted by $\mathbb{D}_{a,s}$:
\begin{equation}
Y_{a,s}=\frac{\mathbb{D}_{a,s-1}\mathbb{D}_{a,s+1}}{\mathbb{D}_{a-1,s}\mathbb{D}_{a+1,s}}
\end{equation} 
where $\mathbb{D}_{a,s}$ satisfies the quantum Hirota equations  eq.(\ref{qHirota}), 
and live in a wider T-hook including the $a=0,2$ and $s=\pm 2$ lines, too. 

\subsubsection{States in the $su(2)$ sector}
Here we present the asymptotic solutions of these equations relevant for 
$Y=0$ brane for states in the $SU(2)$ subsector where the multiparticle states are
composed of particles of $1\dot{1}$. We use the $SU(2)$ grading 
as in this case all the auxiliary $y$ and $w$ roots are absent. 
We analyze the general case afterwards. 

The asymptotic transfer matrices $\mathbb{D}_{a,1}$ are generated by the generating 
functional
\begin{eqnarray}
\mathcal{W}_{su(2)}^{-1}&=&\Bigl(1-\mathcal{D}F\frac{\cR^{(+)+}}{\cR^{(-)+}}
\mathcal{D}\Bigr)\Bigl(1-\mathcal{D}F\mathcal{D}\Bigr)^{-1}\Bigl(1-\mathcal{D}F\frac{u^{+}}{u^{-}}
\mathcal{D}\Bigr)^{-1}\Bigl(1-\mathcal{D}F\frac{u^{+}}{u^{-}}\frac{\cB^{(-)-}}{\cB^{(+)-}}
\mathcal{D}\Bigr) \nonumber \\
&=&\sum_{a}(-1)^{a}\mathcal{D}^{a}\mathbb{D}_{a,1}\mathcal{D}^{a} \,,
\label{eq:WW}
\end{eqnarray} 
where $\mathcal{D}=e^{-\frac{i}{2}\partial_{u}}$, (and therefore $\mathcal{D}
f = f^{-}\mathcal{D}$) and we normalized these transfer matrices  
similarly to the periodic case
\begin{equation}
F=
\sqrt{\frac{Q^{[2]}(u)}{Q^{[-2]}(u)}\frac{u^-}{u^+}}\Bigl(\frac{x^-}{x^+}\Bigr)^{N+1+L}
\Bigl(\frac{\cR^{(-)+}}{\cR^{(+)+}}\Bigr)\prod\limits_{i=1}^N\sigma(p,p_i)\sigma(p_i,-p)\,.
\end{equation}
The functions $\cB^{(\pm)},\cR^{(\pm)}$ are defined as follows
\begin{eqnarray}
&& \cR^{(\pm)} = \prod_{i=1}^{N}\left(x(p)-x^{\mp}(p_{i})\right)
\left(x(p)+x^{\pm}(p_{i})\right)\,, \quad  Q(u)=\prod_{i=1}^{N}(u-u_{i})(u+u_{i})
 \nonumber \\
&& \cB^{(\pm)} = \prod_{i=1}^{N}
\left(\frac{1}{x(p)}-x^{\mp}(p_{i})\right)\left(\frac{1}{x(p)}+
x^{\pm}(p_{i})\right),\quad x^{\pm} +\frac{1}{x^{\pm}} =\frac{1}{g} \left(u\pm \frac{i}{2}\right) \, .
\end{eqnarray}The spectral parameter \(u\) is related to the momentum as \(u=\frac{1}{2}\cot(\frac{p}{2})\sqrt{1+16g^2\sin^2(\frac{p}{2})}\). 

Note that this $N$ particle eigenvalue of the fundamental double row transfer
matrix is similar to the $2N$ particle eigenvalue of the bulk transfer matrix 
where there is a \lq\lq doubling'' of particles: to every particle with $x_j$
there is a \lq\lq reflected'' one with $-x_j$. The presence of the $\frac{u^+}{u^-}$
factors is an extra modification that can be attributed to the boundary. 

The $su(2)$ sector is symmetric $\mathbb{D}_{a,-1}=\mathbb{D}_{a,1}$ and asymptotically we 
have $\mathbb{D}_{a,0}=1$, from which $\mathbb{D}_{a,\pm 2}$ can be calculated by the
equations  (\ref{qHirota}). $Y_{a,0}$ is given by the standard expression
$Y_{a,0}=\frac{\mathbb{D}_{a,1}\mathbb{D}_{a,-1}}{\mathbb{D}_{a+1,0}\mathbb{D}_{a-1,0}}=
\mathbb{D}_{a,1}\mathbb{D}_{a,-1}$. 
In  the following we are interested in the scaling
limit of $\mathbb{D}_{a,s} $, in particular whether it
may be identified with $D_{a,s}$.     

Now we consider the scaling strong coupling limit ($g\rightarrow\infty$) of
the asymptotic large $L$ solution of the $Y=0$ brane's $T$ system described
above. In this limit the length $L$ and the number of particles $N$ (and also,
if present, the number of auxiliary Bethe roots) go to infinity $L\sim N\sim
g$. To describe this limit we introduce a new variable $z$ instead of $u$:
$u=2gz$ such that 
\begin{equation}
x(z)=z+i\sqrt{1-z^2},\qquad x_j=x^{\rm ph}(z_j)=z_j+\sqrt{z_j-1}\sqrt{z_j+1},
\end{equation}
and $x^\pm(z)=x(z\pm\frac{i}{4g})$, $\quad x_j^\pm=x^{\rm
  ph}(z_j\pm\frac{i}{4g})$. Treating $i/(4g)$ as a small parameter, after a
straightforward computation one finds that the strong coupling limit 
of the various functions appearing in  $\mathcal{W}_{su(2)}$ are
\begin{eqnarray}
\frac{\cR^{(+)+}}{\cR^{(-)+}} &\simeq &
  f(z)=\exp\Bigl[\frac{i}{g}(G_-(x)+G_+(x))\Bigr] \quad; \qquad \
G_\mp(x)=\sum\limits_{j=1}^N\frac{x_j^2}{x_j^2-1}\frac{1}{x\mp x_j} \nonumber \\
\frac{\cB^{(-)-}}{\cB^{(+)-}}&\simeq &
  \tilde{f}(z)=\exp\Bigl[-\frac{i}{g}(G_-(1/x)+G_+(1/x))\Bigr], \quad
\frac{u^+}{u^-} \simeq  h(z)=\exp\Bigl(\frac{i}{2gz}\Bigr)
\end{eqnarray} 
and
\begin{equation}
F\simeq 
 \Phi(z)=\exp\Bigl[-\frac{L+1+\sum\limits_{j=1}^NE_j^{(1)}}{2g\sqrt{1-z^2}}-\frac{i}{4gz}\Bigr]\,,
\end{equation}
where, at leading order, $E_j^{(1)}=\frac{x_j^2+1}{x_j^2-1}$ is the energy of
the $j$-th fundamental particle. Here we used the AFS phase for the dressing factor
$\sigma(p,p_i)$ in the strong coupling limit \cite{Arutyunov:2004vx, Gromov:2008ec} which is given as
\begin{eqnarray}
\log \sigma(z,x_{j}) & = &  \log \left( \frac{1-\frac{1}{x^{-}(z) x_{j}^{+}}}
{1-\frac{1}{x^{+}(z) x_{j}^{-}}} \right) +2 i g (z_{j} -z ) \log 
\left(\frac{x^{-}(z)x_{j}^{-} -1}{x^{+}(z)x_{j}^{-} -1} 
\frac{x^{+}(z)x_{j}^{+} -1}{x^{-}(z)x_{j}^{+} -1} \right) \cr
&\simeq & {\frac{i(x(z)-x_{j}) }{g(-1+x(z)^2) (-1+x(z) x_{j}) (-1+x_{j}^2)} } 
\end{eqnarray}
where the mirror variable is denoted by $z(p)$ and one can use $x^{\pm}(-p) = -x^{\mp} (p)$ 
for $\sigma(p_i,-p)$. Also, in the scaling limit the shifted spectral parameters $x^{\pm}$ becomes
\begin{equation}
x^{\pm}(z) = x(z) \pm \frac{i}{2 g} \frac{x^2(z)}{x^2(z) -1} + O(1/g^2),
\end{equation}
as we expand $x^{\pm}(z) \simeq x(z) \pm \frac{i}{4g} \partial_{z} x(z)$.

For the expansion of $\mathcal{W}_{su(2)}^{-1}$ (or $\mathcal{W}_{su(2)}$), 
eq.(\ref{eq:WW}), in the scaling limit it is important to emphasize that in
this limit the operator $\mathcal{D}$ serves only as a formal expansion
parameter since the $\pm i/(4g)$ shifts it generates become negligible. 
The limit of $\mathcal{W}_{su(2)}$ becomes
\begin{equation}
\mathcal{W}_{su(2)}\simeq
\frac{(1-h(z)\Phi(z)\mathcal{D}^2)(1-\Phi(z)\mathcal{D}^2)}{(1-\Phi(z)h(z)\tilde{f}(z)
\mathcal{D}^2)(1-\Phi(z)f(z)\mathcal{D}^2)}\equiv
\tilde{\mathcal{W}}_{su(2)}\,,
\label{qlimit}
\end{equation}
and the scaling limit of ${\mathbb{D}_{a,1} } \simeq  { \tilde  D }_{a,1}  $ is determined
by
$(\tilde{W}_{SU(2)})^{-1}=\sum_{a}(-1)^{a} { \tilde  D }_{a,1}  \mathcal{D}^{2a}$. 
In the classical theory the generating function of the $SU(2,2\vert 4)$
(super)characters of the symmetric representations is
\begin{equation}
w_{4\vert 4}=\frac{(1-y_1t)(1-y_2t)}{(1-x_1t)(1-x_2t)}\times
\frac{(1-y_3t)(1-y_4t)}{(1-x_3t)(1-x_4t)}=\hat{W}^L(x_1,x_2;y_1,y_2)
\hat{W}^R(x_3,x_4;y_3,y_4),
\end{equation}
if $(x_1,\dots x_4\vert y_1,\dots y_4)$ denotes the set of eigenvalues of the
group element and it is described in \cite{Gromov:2010vb} how the characters
$T_{a,s}(x_1,\dots x_4\vert y_1,\dots y_4)$ satisfying
\begin{equation}
T_{a,s}(x_1,\dots x_4\vert y_1,\dots y_4)=T_{a,-s}(1/x_4,\dots 1/x_1\vert
1/y_4,\dots 1/y_1)
\end{equation}
can be obtained from $w_{4\vert 4}$. We want to use this machinery for the
\lq\lq open'' monodromy matrix $T(x)$ to construct $D_{a,s}$ in the
$SU(2)$ subsector. To respect the $D_{a,s}=D_{a,-s}$ symmetry we assume
$\lambda_1=1/\lambda_4$, $\lambda_2=1/\lambda_3$, $\mu_1=1/\mu_4$,
$\mu_2=1/\mu_3$, and write
\begin{equation}
\hat{W}^L=\frac{(1-\mu_1t)(1-\mu_2t)}{(1-\lambda_1t)(1-\lambda_2t)}=\hat{W}^R\, ,
\label{classic}
\end{equation}
with $(\hat{W}^L)^{-1}=\sum_a(-1)^aD_{a,1}t^a$. The key to identify
$\hat{W}^L$ and $\tilde{W}_{SU(2)}$ is to find a relation between the two
formal expansion parameters $t$ and $\mathcal{D}$ that guarantees that
$ \tilde{D}_{a,1} =D_{a,1}$
Following \cite{Gromov:2010vb} to this end we propose the relation
$\mathcal{D}^2=\Phi t$. With this choice one can do even more: by comparing 
(\ref{qlimit}) and (\ref{classic}) one can identify the $\lambda_1$,
$\lambda_2$, $\mu_1$, $\mu_2$ eigenvalues of $T(x)$ with certain
limiting functions
\begin{eqnarray}
\mu_1 &=& \Phi=\exp\Bigl[-\frac{L+1+\sum\limits_{j=1}^NE_j^{(1)}}{2g\sqrt{1-z^2}}-\frac{i}{4gz}\Bigr]
;\quad
\mu_2= h\Phi =
\exp\Bigl[-\frac{L+1+\sum\limits_{j=1}^N
    E_j^{(1)}}{2g\sqrt{1-z^2}}+\frac{i}{4gz}\Bigr]
\nonumber\\
\lambda_1 &=&  \tilde{f}\Phi =
\exp\Bigl[-\frac{J}{2g\sqrt{1-z^2}}-\frac{i}{4gz} - \frac{i}{g}(H_-(1/x)+H_+(1/x))\Bigr]\,,\nonumber\\
\lambda_2 &=&  h f\Phi
=\exp\Bigl[-\frac{J}{2g\sqrt{1-z^2}}+\frac{i}{4gz}+\frac{i}{g}(H_-(x)+H_+(x))\Bigr]
\label{eigenvals}
\end{eqnarray}
where $H_\mp(x)=\sum\limits_{j=1}^N\frac{x^2}{x^2-1}\frac{1}{x\mp x_j}$ and $J=L+1+N$.
We use $\frac{1}{2g\sqrt{1-z^2}}=\frac{i}{g}\frac{x}{x^2-1}$ and
$\sum_jE_j^{(1)}=N+2\sum_j\frac{1}{x_j^2-1}$ (that was also
exploited to obtain (\ref{eigenvals})) in eq.(\ref{qmomenta})  to connect the
eigenvalues of $T(x)$ with the quasi-momenta:
\begin{eqnarray}
\hat{p}_1 (x) &=& -\hat{p}_4 (x) =-\hat{p}_2 (1/x)= \hat{p}_3 (1/x) =
\frac{(J+2\mathcal{Q}_2)x}{g(x^2-1)}+B(x) 
\\
\tilde{p}_1(x) &=&-\tilde{p}_4 (x) =-\tilde{p}_2 (1/x)= \tilde{p}_3 (1/x) =
\frac{Jx}{g(x^2-1)}+B(x)+\frac{1}{g}(H_-(1/x)+H_+(1/x))
\nonumber
\end{eqnarray}
where the boundary contribution is $B(x)=\frac{1}{2g}\frac{x}{x^2+1}$ 
and $\mathcal{Q}_2=\sum_j\frac{1}{x_j^2-1}$. 
It is interesting to compare these quasi-momenta with the ones for the closed
string case presented in \cite{Beisert:2005bm} and in \cite{Gromov:2010vb}:
the (conserved) quantity $\mathcal{Q}_2$ is present in
$\hat{p}_i$ just like in the closed string case,  while the 
$\mathcal{Q}_1=\sum_j\frac{x_j}{x_j^2-1}$ is absent from $\tilde{p}_i$. This
can be understood by recalling the \lq\lq doubling'' of particles mentioned
above: since to every particle with $x_j$ there is another one with $-x_j$ 
that is why $\mathcal{Q}_2$ appears with a factor of $2$ while $\mathcal{Q}_1$ indeed
cancels. Note that this argument also explains why the sums of $H_-$ and $H_+$
with various arguments appear in the quasi-momenta. $N$ (the number of
particles) 
appears in the quasi-momenta as a result of
    working in the $SU(2)$ grading - see the section on the duality
    transformation.  
    Also, one can see that only the quasi-momenta $\tilde{p}_2$ and
    $\tilde{p}_3$, which correspond to $S^3 \subset S^5$, have resolvents
    corresponding to some particle excitations while the total set of quasimomenta
    satisfies all the symmetry and synchronization constraints.
    We also
      note that the boundary contribution, $B(x)$, gives a new pole structure in
      the Riemann surface as a quantum effect.

\subsubsection{Generic states}

Next we turn to the discussion of generic states with $2m_1^L$ ($2m_1^R$)  $y$
roots and $2m_2^L$ ($2m_2^R$) $w$ roots in the $SU(2\vert 2)_L$ ($SU(2\vert
2)_R$)  eigenvalues of the corresponding double row transfer matrices. A
    new feature of the $Y=0$ brane's ABA is that there are only two types of
    auxiliary roots as opposed to the three types present in the closed
    string/bulk case, see \cite{Bajnok:2012xc} and \cite{Galleas:2009ye}.  
To describe these roots we introduce 
\begin{eqnarray}
\cB_{1}^L\cR_{3}^L&=&\prod_{j=1}^{m_{1}^L}\left(x(p)-y_{j}^L\right)\left(x(p)+y_{j}^L\right)\,, \qquad
\cR_{1}^L\cB_{3}^L=\prod_{j=1}^{m_{1}^L}\left(\frac{1}{x(p)}-y_{j}^L\right)
\left(\frac{1}{x(p)}+y_{j}^L\right)\,,
\nonumber \\
Q_{2}^L(u)&=& \prod_{l=1}^{m_{2}^L}(u-w_{l}^L)(u+w_{l}^L)\,.
\end{eqnarray}
The generating functional for the eigenvalues of the $\mathbb{D}_{a,1}$ double
row transfer matrices in antisymmetric representations can be written in terms
of these quantities as \cite{Bajnok:2012xc}
\begin{eqnarray}
(\mathcal{W}_{su(2)}^L)^{-1}&=&\left(1-\mathcal{D}F^L\frac{\cR^{(+)+}}{\cR^{(-)+}}
\frac{\cB_{1}^{L-}\cR_{3}^{L-}}{\cB_{1}^{L+}\cR_{3}^{L+}}\mathcal{D}\right)\left(1-
\mathcal{D}F^L\frac{\cB_{1}^{L-}\cR_{3}^{L-}}{\cB_{1}^{L+}\cR_{3}^{L+}}\frac{Q_{2}^{L++}}{Q_{2}^L}
\mathcal{D}\right)^{-1}\qquad\qquad\qquad 
\\ \nonumber &&
\times\left(1-\mathcal{D}F^L\frac{u^+}{u^-}\frac{\cR_{1}^{L+}\cB_{3}^{L+}}{\cR_{1}^{L-}\cB_{3}^{L-}}
\frac{Q_{2}^{L--}}{Q_{2}^L}\mathcal{D}\right)^{-1}\left(1-\mathcal{D}F^L\frac{u^+}{u^-}
\frac{\cB^{(-)-}}{\cB^{(+)-}}\frac{\cR_{1}^{L+}\cB_{3}^{L+}}{\cR_{1}^{L-}\cB_{3}^{L-}}\mathcal{D}\right) 
\\ \nonumber
&=& \sum_{a} (-1)^{a} \mathcal{D}^a \mathbb{D}_{a,1} \mathcal{D}^a \,,
\end{eqnarray} 
where $su(2)$ refers to the fact that we are working in the $su(2)$ grading while  $\mathbb{D}_{a,-1}$ is obtained by replacing every quantity
here with upper index $L$ with the corresponding quantity with upper index
$R$. 
Here \begin{equation}
F^L=
\sqrt{\frac{Q^{[2]}(u)}{Q^{[-2]}(u)}\frac{u^-}{u^+}}\Bigl(
\frac{x^-}{x^+}\Bigr)^{N-m_1^L+1+L}\Bigl(\frac{\cR^{(-)+}}{\cR^{(+)+}}\Bigr)
\prod\limits_{i=1}^N\sigma(p,p_i)\sigma(p_i,-p)\,,
\end{equation}
and $F^R=F^L(m_1^L\rightarrow m_1^R)$. Thus, the two wings are not symmetric any more,
$\mathbb{D}_{a,s}\ne\mathbb{D}_{a,-s}$, 
but $\mathbb{D}_{a,0}=1$ and $Y_{a,0}=\frac{\mathbb{D}_{a,1}\mathbb{D}_{a,-1}}
{\mathbb{D}_{a+1,0}\mathbb{D}_{a-1,0}}=
\mathbb{D}_{a,1}\mathbb{D}_{a,-1}$.

In the scaling limit one finds  that $F^{L,R}\simeq
\Phi^{L,R}(z)$, where
\begin{equation}
\Phi^L(z)=\exp\Bigl[-\frac{L-m_1^L+1+\sum\limits_{j=1}^NE_j^{(1)}}{2g\sqrt{1-z^2}}
-\frac{i}{4gz}\Bigr]\,,\quad 
\Phi^R(z)=\Phi^L(z)(m_1^L\rightarrow m_1^R). 
\end{equation}  
To describe the scaling limit of the generating functionals
$\mathcal{W}_{su(2)}^{L,R}$ we introduce the
limiting functions
\begin{eqnarray}
\frac{\cR_{1}^{L+}\cB_{3}^{L+}}{\cR_{1}^{L-}\cB_{3}^{L-}}& \simeq &\mathcal{P}^L(z)=
\exp\Bigl(\frac{i}{g}(K^L_-(1/x)+K^L_+(1/x))\Bigr) \quad ;\quad 
K^L_{\pm}(x)=\sum\limits_{i=1}^{m_1^L}\frac{x^2}{x^2-1}\frac{1}{x\pm
  y^L_i} \nonumber \\
\frac{\cB_{1}^{L-}\cR_{3}^{L-}}{\cB_{1}^{L+}\cR_{3}^{L+}}& \simeq &\mathcal{M}^L(z)=
\exp\Bigl(-\frac{i}{g}(K^L_-(x)+K^L_+(x))\Bigr) \nonumber \\
\frac{Q_{2}^{L--}}{Q_{2}^L}&\simeq &
q^L(z)=\exp\Bigl(-\frac{i}{g}(V^L_-(x)+V^L_+(x)+V^L_-(1/x)+V^L_+(1/x))\Bigr)
\nonumber \\
\frac{Q_{2}^{L++}}{Q_{2}^L}& \simeq &
(q^L(z))^{-1} \quad ;\quad V^L_{\pm}(x)=\sum\limits_{l=1}^{m_2^L}\frac{x^2}{x^2-1}\frac{1}{x\pm
  Y^L_l}
\end{eqnarray}
where $Y^L_l$ is defined as 
 $\ Y^L_l=z^L_l+\sqrt{z^L_l-1}\sqrt{z^L_l+1}$, and $w_l^L=2gz^{L}_l$. Using
these functions the limiting expression of the generating functional becomes
\begin{equation}
\mathcal{W}_{su(2)}^L\simeq
\frac{\Bigl(1-h(z)\Phi^L(z)\mathcal{P}^L(z)q^L(z)\mathcal{D}^2\Bigr)\Bigl(1-\Phi^L(z)\mathcal{M}^L(z)(q^L(z))^{-1}
\mathcal{D}^2\Bigr)}{\Bigl(1-h(z)\Phi^L(z)\tilde{f}(z)\mathcal{P}^L(z)\mathcal{D}^2\Bigr)\Bigl(1-f(z)\Phi^L(z)
\mathcal{M}^L(z)\mathcal{D}^2\Bigr)}\equiv
\tilde{\mathcal{W}}_{su(2)}^L\,,
\label{qlimitgen}
\end{equation} 
($\mathcal{W}_{su(2)}^R$ is obtained by replacing every quantity
here with upper index $L$ with the corresponding quantity with upper index
$R$).

Now if we want to use the procedure of \cite{Gromov:2010vb} for $T(x)$
to construct $D_{a,s}$ in the generic case, then, in the lack of the
$D_{a,s}=D_{a,-s}$ symmetry, we write \cite{Gromov:2010vb}
\begin{equation}
\hat{W}^L=\frac{(1-\mu_1t^L)(1-\mu_2t^L)}{(1-\lambda_1t^L)(1-\lambda_2t^L)},\qquad\quad 
\hat{W}^R=\frac{(1-t^R/\mu_4)(1-t^R/\mu_3)}{(1-t^R/\lambda_4)(1-t^R/\lambda_3)}\,,
\label{classicgen}
\end{equation}
where the $t^L$ and $t^R$ formal expansion parameters are related to
$\mathcal{D}^2$: 
\begin{equation}
\mathcal{D}^2=\Phi^Lt^L,\qquad\qquad\mathcal{D}^2=\Phi^Rt^R, 
\label{rel}
\end{equation}
These lead to the following  quasi-momenta 
\begin{eqnarray}
\hat{p}_1(x) = -\hat{p}_2(1/x) &=&
\frac{(J-m_1^L+2\mathcal{Q}_2)x}{g(x^2-1)}+B(x)
-\frac{1}{g}\sum\limits_{\epsilon=\pm}(K^L_\epsilon(1/x)-V^L_\epsilon(x)-V^L_\epsilon(1/x))
\nonumber \\
\hat{p}_3 (x) = -\hat{p}_4 (1/x) &=&
-\frac{(J-m_1^R+2\mathcal{Q}_2)x}{g(x^2-1)}+B(x) 
-\frac{1}{g}\sum\limits_{\epsilon=\pm}(K^R_\epsilon(x)-V^R_\epsilon(x)-V^R_\epsilon(1/x))
 \nonumber \\
\tilde{p}_1(x) = -\tilde{p}_2(1/x)&=&
\frac{(J-m_1^L)x}{g(x^2-1)}+B(x)+\frac{1}{g}
\sum\limits_{\epsilon=\pm}(H_\epsilon(1/x)-K^L_\epsilon(1/x)) \cr
\tilde{p}_3(x) = -\tilde{p}_4(1/x)&=&
-\frac{(J-m_1^R)x}{g(x^2-1)}+B(x)+\frac{1}{g}\sum
\limits_{\epsilon=\pm}(H_\epsilon(x)-K^R_\epsilon(x)) 
\label{rightqm}
\end{eqnarray}
The root configurations represented by the resolvents can condense into cuts in the scaling limit. We will consider square roots and logarithmic cuts in section 6. We also notice that the boundary contribution $B(x) $, is present for any state and  gives new pole structure in the Riemann surface for the spectral curve which can be interpreted as a boundary quantum effect.  

\subsubsection{Duality transformation}

So far our considerations relied on the eigenvalue of the double row transfer
matrix in the $su(2)$ sector. In the literature the $sl(2)$ sector is studied 
most frequently. The eigenvalues in these two sectors are connected by a
duality transformation on the $y$ roots (see Appendix C of
\cite{Bajnok:2012xc}). Here we use this transformation to obtain the
quasi-momenta corresponding to the $sl(2)$ grading.

In this transformation the $2m_1^A$ ($A=L,\, R$) $y$ roots ($y_i^A$ and
$-y_i^A$) are exchanged for $2\tilde{m}_1^A$ dual roots $\tilde{y}$ while the
$w$ roots are not changed ($\tilde{m}_2^A=m_2^A$). The number of dual roots is
determined by the relation 
$\tilde{m}_1^A=N+2m_2^A-m_1^A$ and we introduce the $\tilde{\cal{M}}_A$ and
$\tilde{\cal{P}}_A$ resolvents of the dual $y$ roots in analogy with
${\cal M}_A$ and ${\cal P}_A$. Computing the scaling limit of equations (C.6) and
(C.7) of \cite{Bajnok:2012xc} yields
\begin{equation}
{\cal M}_A=\exp\Bigl( -\frac{i2m_2^A x}{g(x^2-1)}\Bigr)\tilde{\cal
   M}_A^{-1}\exp\Bigl(\frac{i}{g}\sum\limits_{\epsilon=\pm}H_\epsilon(1/x)\Bigr)q_A,
\label{dualm}
\end{equation}
 \begin{equation}
{\cal P}_A=\exp\Bigl( -\frac{i2m_2^A x}{g(x^2-1)}\Bigr)\tilde{\cal
   P}_A^{-1}\exp\Bigl(-\frac{i}{g}\sum\limits_{\epsilon=\pm}H_\epsilon(x)\Bigr)q_A^{-1},
\label{dualp}
\end{equation}  
while the dual version of $\Phi^A (z)$ is obtained as $\tilde{\Phi}^A(z)=
\Phi^A(z)\ [m_1^A\rightarrow N+2m_2^A-\tilde{m}_1^A]$. The $\lambda_i$, $\mu_i$
eigenvalues in the $sl(2)$ grading are obtained from the previous ones in the
$su(2)$ grading by
replacing $\Phi^A$ with $\tilde{\Phi}^A$ and also substituting
eq.(\ref{dualm}) and eq.(\ref{dualp}).

The simplest situation is when we consider $N$ fundamental particles of
$3\dot{3}$ type with no auxiliary (dual) roots $\tilde{m}^A_1=0=m_2^A$. Note
that this requires a non vanishing $m_1^A$, in fact $m_1^A=N$, but
eq.s(\ref{dualm}-\ref{dualp}) simplify and eventually one finds the quasi-momenta    
\begin{eqnarray}
\hat{p}_1 (x) &=& -\hat{p}_2 (1/x)=-\hat{p}_4(x)=\hat{p}_3(1/x)=
\frac{(L+1+2\sum_j\frac{1}{x_j^2-1})x}{g(x^2-1)}+B(x)+\frac{1}{g}\sum
\limits_{\epsilon=\pm}H_\epsilon(1/x)
\nonumber\\
\tilde{p}_1(x) &=& -\tilde{p}_2(1/x)=-\tilde{p}_4(x) =\tilde{p}_3(1/x)=
\frac{(L+1)x}{g(x^2-1)}+B(x)
\end{eqnarray}
Note that $N$, the number of particles, disappeared from the quasimomenta, and
also the non vanishing resolvent densities moved from the $S^5$ components of
the quasimomenta to the $AdS_5$ ones.  

\section{Quasimomenta from the all-loop boundary Bethe equations}

In this section we derive the quasi-momenta (\ref{rightqm})
from the scaling limit of the all-loop boundary Bethe equations which were constructed for the $Y=0$ brane set-up in 
\cite{Galleas:2009ye,Nepomechie:2009zi}. We start with the $su(2)$ sector. 

\subsection{$su(2)$ sector}

In the $su(2)$ sector we consider $N$ particles, with rapidities $u_j$ but without any polarization, 
in a finite volume $L$, satisfying the $Y=0$ brane boundary conditions on both ends. 
The only Bethe Ansatz equation in terms of $Y_{1,0}$ function reads \(\)as 
\begin{equation}
Y_{1,0}(u_j)=\mathbb{D}_{1,1}(u_j)\mathbb{D}_{1,-1}(u_j)=\Bigl(\frac{x_{j}^-}{x_{j}^+}\Bigr)^{2L}
\frac{Q^{[2]}(u_{j})u^{-}_{j }}{Q^{[-2]}(u_j)u^{+}_{j}}\prod\limits_{i=1}^N\sigma^{2}(p_{j},p_i)\sigma^{2}(p_i,-p_{j})\,=-1
\label{YBA}
\end{equation}
If we take logarithm of (\ref{YBA}) with $g \sim L \sim N \gg 1$, the scaling Bethe equation becomes 
\begin{equation}
2 \pi n = \frac{2J x}{g(x^2-1)}-2B(x) -\frac{2}{g}\sum_{\epsilon=\pm} 
H_{\epsilon}(x) 
\end{equation}
where we used the boundary contribution $B(x)$ and defined $J= L+ N$. 
In the $su(2)$ sector there is two-sheeted Riemann surface corresponding to $p(x)$ and $-p(x)$. These sheets are connected through  cuts where  Bethe roots with given mode numbers condense.   The   Bethe equation relates the quasimomenta \begin{equation}
p(x)= \frac{J x}{g(x^2-1)}-B(x)-\frac{1}{g}\sum\limits_{\epsilon=\pm} H_\epsilon(x)
\end{equation}on the two sides of the cut  as
\begin{equation}
p(x + i 0) + p(x - i 0) = 2 \pi n   
\end{equation}
where $x$ belongs to the cut joining two sheets with mode number $n$.

\subsection{Generic case} 

In the generic case we have not only the massive Bethe equation (\ref{YBA}) but also the 
magnonic ones coming from the regularity of the double row transfer matrix, $\mathbb{D}_{1,1}$,  at the 
auxiliary root positions:
\begin{equation} 
\frac{\cR^{(+)+}Q_2^L}{\cR^{(-)+}Q_2^{L++}} \biggr|_{x^+=\pm y_j^L} =1=  \frac{\cB^{(-)-}Q_2^L}{\cB^{(+)-}Q_2^{L--}} \biggr|_{x^-=\pm 1/y_j^L}
;\quad \frac{Q_{1}^{L-}Q_{3}^{L-}Q_2^{++}u^-}{Q_{1}^{L+}Q_{3}^{L+}Q_2^{--}u^+} \biggr|_{u=\pm w_j^L}=-1
\end{equation}
It is more natural to rewrite these equations into a manifestly $\mathrm{psu}(2,2\vert 4)$ covariant way. 
In so doing we relabel the roots as
\begin{eqnarray}
x_{j} & \longleftrightarrow & x_{4,j}\; , \quad  y_{j}^{L}  \longleftrightarrow  \frac{1}{x_{1,j}}\; , 
\quad y_{K_{1} +j}^L  \longleftrightarrow  x_{3,j}\; , \quad w_{j}^{L}  \longleftrightarrow  x_{2,j} 
\; ,\quad K_{2}\longleftrightarrow m_2^{L} \nonumber \\ N & \longleftrightarrow & K_4 \; , \quad
y_{j}^{R}  \longleftrightarrow  x_{5,j}\; , \quad  y_{K_{5} +j}^R  \longleftrightarrow  \frac{1}{x_{7,j}} 
\; , \quad
w_{j}^{R}  \longleftrightarrow  x_{6,j}\; ,\quad  K_{6}\longleftrightarrow m_{2}^{R} 
\end{eqnarray}
where we split the roots $y^{L/R}_j$ according to their absolute value, thus $m_1^{L} = K_1 + K_3$
and $m_1^{R} = K_5 + K_7$. 
In the following we analyze the scaling limit
\begin{equation}
g\sim u_{a} \sim K_{a} \sim L \gg 1, \quad a=1,2,...,7,
\end{equation}

In this scaling limit, using the Riemann surfaces structure of the closed
string,  
the Bethe equations can be written as: 
\footnote{Actually, the existence of the  $Y=0$ giant graviton breaks the residual symmetry $SU(2|2)^2$ to 
$SU(1|2)^2$ ~by the term \(B(x)\).    }
\begin{eqnarray}
2 \pi n_{{\tilde 2}{\tilde 3}} &=& \frac{2J x}{g(x^2-1)}-2B(x) -\frac{1}{g}\sum_{\epsilon=\pm} 
\left(2 H_{\epsilon}^4(x) - {H}_{\epsilon}^7 (1/x) -{H}_{\epsilon}^1 (1/x) 
- H_{\epsilon}^3 (x) - H_{\epsilon}^5 (x) \right) \cr
2 \pi n_{{\hat 1}{\tilde 1}} &=& -\frac{2 G_{4}'(0) x}{g(x^2-1)} +\frac{1}{g}\sum_{\epsilon=\pm} 
\left( H_{\epsilon}^2 (x) + H_{\epsilon}^2 (1/x) - H_{\epsilon}^4 (1/x)\right) \cr
2 \pi n_{{\tilde 4}{\hat 4}} &=& -\frac{2 G_{4}'(0) x}{g(x^2-1)} +\frac{1}{g}\sum_{\epsilon=\pm} 
\left( H_{\epsilon}^6 (x) + H_{\epsilon}^6 (1/x) - H_{\epsilon}^4 (1/x)\right) \cr
2 \pi n_{{\tilde 2}{\hat 2}} &=& \frac{2 G_{4}'(0) x}{g(x^2-1)} -\frac{1}{g}\sum_{\epsilon=\pm} 
\left( {H}_{\epsilon}^4 (x) - H_{\epsilon}^2 (x) - H_{\epsilon}^2 (1/x)\right)\label{scalingBethe}\\
2 \pi n_{{\hat 3}{\tilde 3}} &=& \frac{2 G_{4}'(0) x}{g(x^2-1)} -\frac{1}{g}\sum_{\epsilon=\pm} 
\left( {H}_{\epsilon}^4 (x) + H_{\epsilon}^6 (x) + {H}_{\epsilon}^6 (1/x) \right) \cr
2 \pi n_{{\hat 1}{\hat 2}} &=& 2B(x)+\frac{1}{g}\sum_{\epsilon=\pm} \left(2 H_{\epsilon}^2 (x) 
+ 2 H_{\epsilon}^2 (1/x) - H_{\epsilon}^1 (x)- H_{\epsilon}^1 (1/x) 
- H_{\epsilon}^3 (x) - H_{\epsilon}^3 (1/x)\right) \cr
2 \pi n_{{\hat 3}{\hat 4}} &=& 2B(x)+\frac{1}{g}\sum_{\epsilon=\pm} \left(2 H_{\epsilon}^6 (x) 
+ 2 H_{\epsilon}^6 (1/x) - H_{\epsilon}^5 (x)- H_{\epsilon}^5 (1/x) 
- H_{\epsilon}^7 (x) - H_{\epsilon}^7 (1/x)\right)\nonumber
\end{eqnarray}
where we used $H_{\pm}^{i}(x)=\sum\limits_{j=1}^{K_{i}} \frac{x^2}{x^2 -1} \frac{1}{x \pm x_{i,j}} $, 
$G_{4}(x)=\sum\limits_{j=1}^{K_{4}} \frac{x_{4,j}^2}{x_{4,j}^2 -1} \frac{1}{x \pm x_{4,j}} $ and
defined $J$ as $J=L + K_{4} + \frac{K_{1}-K_{3}+K_{7}-K_{5}}{2} +1$.
These Bethe equations correspond to differences between the various quasi-momenta:
\begin{eqnarray}
\hat{p}_1 (x) = -\hat{p}_2(1/x) &=&
+\frac{(J_L +2Q_2) x}{g(x^2-1)}+B(x)-\frac{1}{g}\sum\limits_{\epsilon=\pm}\left( H_\epsilon ^1 (x) 
- H_\epsilon ^2(x) - H_\epsilon ^2(1/x)+H_\epsilon ^3 (1/x) \right) \cr
\hat{p}_3 (x) = -\hat{p}_4(1/x) &=&
-\frac{(J_R +2Q_2) x}{g(x^2-1)}+B(x)-\frac{1}{g}\sum\limits_{\epsilon=\pm}\left(H_\epsilon ^5 (x) 
- H_\epsilon ^6(x)- H_\epsilon ^6(1/x) + H_\epsilon ^7(1/x)\right)
\nonumber\\
\tilde{p}_1 (x) = -\tilde{p}_2 (1/x) &=&
+\frac{J_L x}{g(x^2-1)}+B(x)-\frac{1}{g}\sum\limits_{\epsilon=\pm}\left( H_\epsilon ^1 (x) 
+ H_\epsilon ^3(1/x) - H_\epsilon ^4(1/x)\right) \cr
\tilde{p}_3 (x) = -\tilde{p}_4 (1/x)&=&
-\frac{J_R x}{g(x^2-1)}+B(x)+\frac{1}{g}\sum\limits_{\epsilon=\pm}\left(H_\epsilon ^4 (x) 
- H_\epsilon ^5(x)- H_\epsilon ^7(1/x)\right)
\label{bdryqm}
\end{eqnarray}
where we defined $Q_2 = - G_4'(0)$, $J_L=L+1+K_{4} + K_1 -K_3$ and $J_R=L+1+K_{4} - K_5 +K_7$ 
such that $J_L + J_R = 2 J$. We also used
that \begin{equation}
\sum_{j=1}^{ {K}_{i}} \frac{ x^2 }{ x^2 -1} \frac{1}{x \pm \frac{1}{x_{i,j} }} = 
\frac{ K_{i} x}{x^2-1} + H_{\pm}^{i} (1/x). 
\end{equation}
Then, one can easily check that specific differences between two quasi-momenta give the all-loop boundary 
Bethe equations in the scaling limit (\ref{scalingBethe}). Note that quasi-momenta 
for left wings and for right wings have different dependence on the numbers of Bethe roots. 

We explained the analytic properties of quasimomenta for open strings attached to $Y=0$ brane in section 2. 
Such analytic properties are related to physical information on conserved charges. 
Here, we can read off the same properties from the quasimomenta.
First, let us investigate the synchronization of the residues at $x= \pm 1$. For example, 
the residues of ${\hat p}_{1,2}$ at $x=1$ become
\begin{equation}
{\hat p}_{1,2} \simeq +\frac{(J_L +2Q_2) }{2g}-\frac{1}{2 g} \left( \sum\limits_{j=1}^{K_1} 
\left(\frac{1}{1-x_{1,j}}+\frac{1}{1+x_{1,j}} \right) - 
\sum\limits_{j=1}^{K_3} \left(\frac{1}{1-x_{3,j}}+\frac{1}{1+x_{3,j}} \right)\right), \nonumber
\end{equation}
which is equivalent to that of ${\hat p}_{3,4}$ at $x=1$ 
\begin{equation}
{\hat p}_{3,4} \simeq -\frac{(J_R +2Q_2) }{2g}-\frac{1}{2 g} \left( \sum\limits_{j=1}^{K_5} 
\left(\frac{1}{1-x_{5,j}}+\frac{1}{1+x_{5,j}} \right) - \sum\limits_{j=1}^{K_7} 
\left(\frac{1}{1-x_{7,j}}+\frac{1}{1+x_{7,j}} \right)\right), \nonumber
\end{equation}
as a result of (\ref{scalingBethe}). 
Also, the inversion symmetry between each quasi-momenta and the reflection symmetry can be easily 
checked. Note that the absence of winding in the inversion symmetry corresponds to 
the absence of the 1st conserved charge $Q_1$ in the Bethe equation. Also, the reflection symmetry 
is expected as our quasi-momenta have the doubling nature.\footnote{Therefore, symmetric 
distributions of Bethe roots on complex plane are suitable solutions for Bethe equation.}
Last but not least, the large $x$ asymptotics of the quasi-momenta are given
in terms of the conserved charges as (for details see Appendix \ref{sec:Dynkin}): 
\begin{equation}
\lim_{x\rightarrow\infty}\left(\begin{array}{cc}
{\hat p}_{1}(x) & \,\ {\hat p}_{2}(x) \\
{\hat p}_{3}(x) & \,\ {\hat p}_{4}(x) \\
{\tilde p}_{1}(x) & \,\ {\tilde p}_{2}(x) \\
{\tilde p}_{3}(x) & \,\ {\tilde p}_{4}(x)\\
\end{array}\right)\simeq\frac{1}{g x}\left(\begin{array}{cccc}
\Delta -S_1 +S_2 & \,\ \Delta +S_1 -S_2 \\ 
-\Delta -S_1 -S_2 & \,\ -\Delta +S_1 +S_2\\
J_1 +J_2 -J_3 & \,\ J_1 -J_2 +J_3 \\ 
-J_1 +J_2 +J_3 & \,\ -J_1 -J_2 -J_3
\end{array}\right)\label{largexasymp}
\end{equation}
Compared with the closed string case,  we observe that only the prefactor  has doubled, i.e. each charge comes with a factor two. But, it doesn't mean that open strings have doubled charges:
all doubling nature is just the effect of the double monodromy matrix and the definition of the quasimomenta. Observe that due to the \(Y=0  \) brane boundary condition  we still 
had three angular momenta $J_{1,2,3}$ ,   spins $S_{1,2}$ and  energy $\Delta$   
 coming from the $S^5$ and $AdS_5$ isometries, like in the closed strings
 case, ie. without any $D$-branes.

\section{Quasiclassical fluctuations of open string solutions} 
One advantage of the algebraic curve formalism is that one can efficiently compute semiclassical contributions to  conserved charges from the quasimomenta by exploiting their  analytic properties.
In this section, we will treat two kinds of open string solutions - the BMN string and the boundary giant magnon.

\subsection{BMN string}
Having constructed the quasimomenta for some simple classical string solutions
we 
want to consider the quantum fluctuations around them. This also makes
possible to compare the quasi momenta obtained from the $Y$ system/ABA and the
ones  that describe the quasiclassical fluctuations. 
 To this end we follow the
procedure summarized for the closed string case in \cite{Gromov:2009zz}. In
terms of the algebraic curve it means that we add some microscopic cuts -
i.e. some finite number of poles - to the quasimomenta of the classical
solution. These additional pole terms must satisfy several requirements that
follow from the general equations for the cuts on the Riemann surface.
These requirements fix their form completely and also determine the shift in
the energy corresponding to the fluctuations. We focus mainly on the points in
the procedure that are different from the closed string case.

The quasimomenta for the BMN string are given in (\ref{BMNmomenta}) 
and to describe the quantum fluctuations in all components we make the
substitution $p(x)\rightarrow p(x)+\delta p(x)$. If the microscopic cut
(i.e. the pole) is shared by the sheets $i$ and $j$ then its location
$x^{ij}_n$ is determined in leading order  by the equation
\begin{equation}
p_i(x^{ij}_n)-p_j(x^{ij}_n)=2\pi n, \qquad \vert x^{ij}_n\vert >1\,,
\label{microcut}\end{equation}
and we denote by $N^{ij}_n$ the number of excitations with mode number $n$
between $i$ and $j$ (we also define $N^{ij}=\sum\limits_n N^{ij}_n$). 
In our case the non vanishing $x^{ij}_n$-s are independent of $i,j$ and depend
only on the mode number $n$
\begin{equation}
x^{ij}_n\rightarrow x_n=\frac{1}{n}(\nu +\sqrt{n^2+\nu^2}).
\end{equation}
The quasimomenta $p(x)+\delta p(x)$ should be analytical on the $x$ plane and 
satisfy the following requirements
\begin{itemize}
\item must have poles at $x^{ij}_n$ with residua $\pm\frac{1}{g}\alpha
  (x^{ij}_n)N^{ij}_n$ (where $\alpha (x)=\frac{x^2}{x^2 -1}$).
\item obeying the $x\rightarrow 1/x$ (inversion) and the $x\rightarrow -x$ 
(reflection) symmetry
  properties 
\begin{equation}
\tilde{p}_{1,2}(x)=-\tilde{p}_{2,1}(1/x),\qquad 
\tilde{p}_{3,4}(x)=-\tilde{p}_{4,3}(1/x),\qquad 
\tilde{p}_{i}(-x)=-\tilde{p}_{i}(x),\quad i=1,\dots ,4\,,
\end{equation}
\begin{equation}
\hat{p}_{1,2}(x)=-\hat{p}_{2,1}(1/x),\qquad 
\hat{p}_{3,4}(x)=-\hat{p}_{4,3}(1/x),\qquad 
\hat{p}_{i}(-x)=-\hat{p}_{i}(x),\quad i=1,\dots ,4\,. 
\end{equation}
\item the residua at $x=\pm 1$ should coincide for
  $\hat{p}_1,\hat{p}_2,\tilde{p}_1,\tilde{p}_2$ and for 
$\hat{p}_3,\hat{p}_4,\tilde{p}_3,\tilde{p}_4$ 
\item introducing the notation $\sum\limits_i\equiv
  \sum\limits_{i=\hat{3}\hat{4}\tilde{3}\tilde{4}}$ and 
  $\sum\limits_k\equiv
  \sum\limits_{k=\hat{1}\hat{2}\tilde{1}\tilde{2}}$ the large $x$ asymptotics
  of $\delta p(x)$ should be given by
\begin{equation}
\begin{pmatrix}
\delta\hat{p}_1\\
 \delta\hat{p}_2\\
\delta\hat{p}_3\\
 \delta\hat{p}_4
\end{pmatrix}\sim \frac{1}{xg}\begin{pmatrix}
                          {\delta\Delta}+2\sum\limits_i N^{\hat{1}i}\\
                          {\delta\Delta}+2\sum\limits_i N^{\hat{2}i}\\
                           -{\delta\Delta}-2\sum\limits_k N^{k\hat{3}}\\
                          -{\delta\Delta}-2\sum\limits_k
                          N^{k\hat{4}}
                          \end{pmatrix} \qquad\qquad\begin{pmatrix}
\delta\tilde{p}_1\\
 \delta\tilde{p}_2\\
\delta\tilde{p}_3\\
 \delta\tilde{p}_4
\end{pmatrix}\sim \frac{1}{xg}\begin{pmatrix}
                           -2\sum\limits_i N^{\tilde{1}i}\\
                          -2\sum\limits_i N^{\tilde{2}i}\\
                           2\sum\limits_k N^{k\tilde{3}}\\
                          2\sum\limits_k N^{k\tilde{4}}\end{pmatrix}
\label{deltaasym}\end{equation}
\end{itemize}
The $x\rightarrow -x$ symmetry properties, the absence of winding in
$\tilde{p}$ and the appearance of the factors of two in front of the sums in
the asymptotic expressions are the new features of these requirements when
compared to the closed string case, while $\delta\Delta$ determines in the same
way the energy 
\begin{equation}
E=\delta\Delta +\mathrm{excitation\ numbers}
\end{equation}

The $\delta p_i$ for the BMN case are obtained by enforcing the $x\rightarrow
-x$ symmetry i.e. by writing 
\begin{equation}
g\cdot\delta\hat{p}_2(x)= \hat{\alpha}\frac{2x}{x^2
  -1}+\sum\limits_{i,n}\Bigl(\frac{\alpha
(x^{\hat{2}i}_n)N^{\hat{2}i}_n}{x- x^{\hat{2}i}_n}+\frac{\alpha
(x^{\hat{2}i}_n)N^{\hat{2}i}_n}{x+ x^{\hat{2}i}_n}+\frac{\alpha
(x^{\hat{1}i}_n)N^{\hat{1}i}_n}{1/x- x^{\hat{1}i}_n}+\frac{\alpha
(x^{\hat{1}i}_n)N^{\hat{1}i}_n}{1/x+ x^{\hat{1}i}_n}\Bigr) 
\end{equation}
\begin{equation}
g\cdot\delta\hat{p}_3(x)= \hat{\beta}\frac{2x}{x^2
  -1}+\sum\limits_{k,n}\Bigl(\frac{\alpha
(x^{\hat{3}k}_n)N^{\hat{3}k}_n}{x- x^{\hat{3}k}_n}+\frac{\alpha
(x^{\hat{3}k}_n)N^{\hat{3}k}_n}{x+ x^{\hat{3}k}_n}+\frac{\alpha
(x^{\hat{4}k}_n)N^{\hat{4}k}_n}{1/x- x^{\hat{4}k}_n}+\frac{\alpha
(x^{\hat{4}k}_n)N^{\hat{4}k}_n}{1/x+ x^{\hat{4}k}_n}\Bigr) 
\end{equation}
while $\delta\tilde{p}_2$ is the same as $\delta\hat{p}_2$ with the
substitutions $\hat{1}\hat{2}\rightarrow\tilde{1}\tilde{2}$ plus changing the
signs in front of the sums ($\delta\tilde{p}_3$ is also obtained from 
$\delta\hat{p}_3$ by the substitutions
$\hat{3}\hat{4}\rightarrow\tilde{3}\tilde{4}$ and changing the signs of the
sums). The additional components of $\delta p$ are obtained by exploiting the
inversion symmetry: $\delta\hat{p}_1(x)=-\delta\hat{p}_2(1/x)$, etc. These
expressions reveal several interesting features: they contain only two unknown
parameters ($\hat{\alpha}$ and $\hat{\beta}$) since the reflection symmetry
allows no $x$ independent constant terms. This symmetry also doubled the
poles:  
 to every pole at $x^{ij}_n$ there is another one at  $-x^{ij}_n$, furthermore
 the residua of these poles must be the same. The last sums in these
 expressions (and especially their signs) are introduced to guarantee that
 after exploiting the inversion symmetry we obtain the expected pole terms.   
The appearance of poles at $-x^{ij}_n$ in the Ansatz for $\delta p_i$ may be
understood also by realising that they also solve eq.(\ref{microcut}) but with 
$n\rightarrow -n$ on the right hand side. This means that together with the
microscopic cut corresponding to the integer $n$ in (\ref{microcut}) we also
have a cut corresponding to $-n$, i.e. reflection symmetry doubles the cuts
(in a similar way as inversion symmetry does it).  

Matching the asymptotic behaviour of these $\delta p_i$-s to the one in 
(\ref{deltaasym}) 
determines all the unknown parameters. Indeed from the asymptotics of
$\delta\hat{p}_1/\delta\hat{p}_2$ (respectively
$\delta\hat{p}_3/\delta\hat{p}_4$) we find 
\begin{equation}
{\delta\Delta}=2\hat{\alpha}+\sum\limits_{n,i}\frac{\sqrt{\nu^2+n^2}-\nu}{\nu}(N^{\hat{1}i}_n+N^{\hat{2}i}_n),
\quad-{\delta\Delta}=2\hat{\beta}-\sum\limits_{n,k}\frac{\sqrt{\nu^2+n^2}-\nu}{\nu}(N^{\hat{3}k}_n+N^{\hat{4}k}_n),
\end{equation}  
while from the asymptotics of
$\delta\tilde{p}_1/\delta\tilde{p}_2$ (respectively
$\delta\tilde{p}_3/\delta\tilde{p}_4$) it follows that
\begin{equation}
0=2\hat{\alpha}-\sum\limits_{n,i}\frac{\sqrt{\nu^2+n^2}-\nu}{\nu}(N^{\tilde{1}i}_n+N^{\tilde{2}i}_n),\qquad
0=2\hat{\beta}+\sum\limits_{n,k}\frac{\sqrt{\nu^2+n^2}-\nu}{\nu}(N^{\tilde{3}k}_n+N^{\tilde{4}k}_n).
\end{equation}
A remarkable property of these equations that they determine $\delta\Delta$, 
$\hat{\alpha}$ and $\hat{\beta}$ without imposing any condition on the
excitation numbers $N^{ij}_n$ since by their definition 
$\sum\limits_i(N^{\hat{1}i}_n+N^{\hat{2}i}_n+N^{\tilde{1}i}_n+N^{\tilde{2}i}_n)=
\sum\limits_k(N^{\hat{3}k}_n+N^{\hat{4}k}_n+N^{\tilde{3}k}_n+N^{\tilde{4}k}_n)$.   
We emphasize this because the analogous
equations in the closed string case have a solution only if the excitation
numbers satisfy a condition, which turns out to be the level matching
condition. For open strings there is no level matching condition thus on
physical grounds  we expect that there is also no condition on the excitation
numbers. 
The consistent expression for $\delta\Delta$ is
\begin{equation}
\delta\Delta=\sum\limits_n\frac{\sqrt{\nu^2+n^2}-\nu}{\nu}\sum\limits_i(N^{\hat{1}i}_n+N^{\hat{2}i}_n+N^{\tilde{1}i}_n+N^{\tilde{2}i}_n).
\end{equation}

\subsection{Boundary giant magnon} 
In the bulk the classical giant magnon solution corresponds to a logarithmic cut in complex plane of Bethe roots with $H(x) = - i \log\frac{x-X^{+}}{x-X^{-}}$ \cite{Minahan:2006bd}. As the algebraic curve description for open string is built on  symmetric roots configurations, we propose the quasi-momenta for the classical boundary giant magnon as
\begin{eqnarray}
&& {\hat p}_1 = {\hat p}_2 = -{\hat p}_3 = -{\hat p}_4 = \frac{2 \Delta}{g}\frac{x}{x^2-1} ,  \cr
&& {\tilde p}_2 = \frac{2 \Delta}{g}\frac{x}{x^2-1} - i \log\frac{x-X^{+}}{x-X^{-}} + i \log\frac{x+X^{+}}{x+X^{-}} = - {\tilde p}_3 ,\label{bgm} \\
&& {\tilde p}_1 = \frac{2 \Delta}{g}\frac{x}{x^2-1}- i \log\frac{x- 1/X^{-}}{x- 1/X^{+}} + i \log\frac{x+ 1/X^{-}}{x+ 1/X^{+}}  = - {\tilde p}_4 , \nonumber
\end{eqnarray}
where we replaced the doubled resolvent $H^{4}_{\pm}(x)$ in the generic
quasimomenta (\ref{bdryqm}) with two symmetric logarithmic cuts between $\pm
X^{+}$ and $\pm X^{-}$. Please note that we don't introduce any twist factor
in the quasi-momenta unlike in the periodic case.\footnote{We omitted the boundary contribution $B(x)$ as it's rapidly suppressed in the exponential part of (\ref{oneloope}). However, $B(x)$ might be important in the subleading quantum corrections.}
Then, by (\ref{bgm}), the inversion and reflection symmetries between the quasi-momenta are automatically satisfied.
The dispersion relation of the boundary giant magnon is obtained by the large $x$
asymptotics (\ref{largexasymp}) and $e^{i p} = \frac{X^{+}}{X^{-}}$ and is
given explicitly as \footnote{We'll confine to the simple boundary giant magnon with $J_2 = 1$ and $L=J_1$.}
\begin{equation}
\Delta - J_1 = \sqrt{J_2^2 + 16 g^2 \sin^2{\frac{p}{2}}} \equiv \epsilon(p).
\end{equation} 
Now, let us compute the semi-classical correction for the boundary giant magnon from the algebraic curve.
The quasimomenta for boundary giant magnon can be thought as that of a periodic two-magnon state with the following constraint :
\begin{equation}
X_2^{\pm} = - X_1^{\mp} \equiv - X^{\mp} \label{constraint}
\end{equation}
where the twist factors are naturally cancelled  since $p_1 = -p_2$.
Then, we can take the multi-magnon computation by Hatsuda and Suzuki \cite{Hatsuda:2008na} and carefully impose the constraint (\ref{constraint}). 
All $\delta p_{\hat i}$ would be equivalent to those of giant magnon with the
periodic boundary condition. Only the $\delta p_{\tilde i}$ are  
different because we have to consider the effects of the simple poles at $x=-X^{\pm}$ and $x=-1/X^{\pm}$. For example, one can express $\delta p_{\tilde 1}$ and $\delta p_{\tilde 3}$ as
\begin{eqnarray}
\delta p_{\tilde 1} &=& \frac{A x + B}{x^2 -1} - \sum_{n, j={\tilde 3} {\tilde 4} {\hat 3} {\hat 4}} \left[\frac{N_{n}^{{\tilde 1} j} \alpha(x_{n}^{{\tilde 1} j})}{x- x_{n}^{{\tilde 1} j}} - \frac{N_{n}^{{\tilde 2} j} \alpha(x_{n}^{{\tilde 2} j})}{1/x- x_{n}^{{\tilde 2} j}}   +  \frac{N_{n}^{{\tilde 1} j} \alpha(x_{n}^{{\tilde 1} j})}{x+ x_{n}^{{\tilde 1} j}} - \frac{N_{n}^{{\tilde 2} j} \alpha(x_{n}^{{\tilde 2} j})}{1/x+ x_{n}^{{\tilde 2} j}} \right] \cr
&+& \sum_{\beta=\pm} \left( \frac{A^{\beta}}{X^{\beta} - 1/x} - \frac{A^{\beta}}{X^{\beta} + 1/x}\right), \cr
\delta p_{\tilde 3} &=& -\frac{C x+D}{x^2 -1} + \sum_{n, j={\tilde 1} {\tilde 2} {\hat 1} {\hat 2}} \left[\frac{N_{n}^{ j{ \tilde 3} } \alpha(x_{n}^{j{ \tilde 3} })}{x- x_{n}^{j{ \tilde 2} }} - \frac{N_{n}^{j{\tilde 4} } \alpha(x_{n}^{j{\tilde 4}})}{1/x- x_{n}^{j{\tilde 1}}} + \frac{N_{n}^{ j{ \tilde 3} } \alpha(x_{n}^{j{ \tilde 3} })}{x+ x_{n}^{j{ \tilde 2} }} - \frac{N_{n}^{j{\tilde 4} } \alpha(x_{n}^{j{\tilde 4}})}{1/x+ x_{n}^{j{\tilde 1}}} \right] \cr
&-& \sum_{\beta=\pm} \left( \frac{A^{\beta}}{x-X^{\beta}} + \frac{A^{\beta}}{x+ X^{\beta}}\right), \nonumber
\end{eqnarray}
with unknown $A$, $B$, $C$, $D$ and $A^\beta$. 
Then, the reflection symmetry yields $B=D=0$ and the inversion symmetries between
quasimomenta and the large $x$ asymptotic conditions give a set of equations between the unknown coefficients as in \cite{Hatsuda:2008na}. 
In the periodic case  
one cannot determine the exact form of fluctuation frequencies of
multi-magnon states within the spectral curve method, nevertheless, one can exactly
evaluate the one-loop correction to the energy 
by using the saddle point approximation \cite{Hatsuda:2008na, Ahn:2010eg}.
This happens for the case of boundary giant magnon, too.
We claim that the leading quantum correction to the energy of the boundary giant magnon has the following form: 
\begin{equation}
\delta \epsilon_{\rm 1-loop} = \int \frac{dx}{2\pi i} \partial_{x}\Omega\left(x\right) \sum_{(ij)} \left(-1\right)^{F_{ij}}
e^{-i\left(p_{i}-p_{j}\right)} \label{quaneff}
\end{equation} 
Here, $(i j)$ means all polarization pairs and the fluctuation frequency $\Omega\left(x\right)$ is given as
\begin{equation}
\Omega\left(x\right) = \frac{2}{{x}^2 -1} \left(1 - \frac{X^{-} + X^{+}}{X^{-} X^{+} +1} x\right), \label{flucfre}
\end{equation}
where we used the symmetric additional poles like $x=x_n^{i j}$ near $X_1^{\pm}$ and $x=-x_n^{i j}$ near $X_2^{\pm}$ and it corresponds to $\alpha_{1}=\alpha_{2}=\frac{1}{2}$ in  the notation of \cite{Hatsuda:2008na}.

Then, one can compare (\ref{quaneff}) to the boundary L\"uscher's $F$-term formula to check the semiclassical result. The boundary L\"uscher's $F$-term formula is given by
\begin{equation}
\delta E^{\rm{F}}=-\int_{0}^{\frac{\omega_{1}}{2}}\frac{dz}{2\pi}(\partial_{z}\tilde{p}(z))\mathbb{S}_{ia}^{jb}(\frac{\omega}{2}+z,u)\mathbb{R}_{j}^{k}(\frac{\omega}{2}+z)\mathbb{S}_{lb}^{ka}(\frac{\omega}{2}-z,u)\mathbb{C}^{l\bar{l}}\mathbb{R}_{\bar{l}}^{\bar{i}}(\frac{\omega}{2}-z)\mathbb{C}_{\bar{i}i}e^{-2\tilde{\epsilon}L}, \label{lus}
\end{equation}
where we used the expression of \cite{Bajnok:2010ui} for fundamental virtual particle with $Q =1$ since all other mirror boundstate contributions with $Q>1$ are suppressed in the strong coupling limit.\footnote{$\mathbb{C}$ is the charge conjugation matrix. }
One can rewrite (\ref{lus}) to a form  more appropriate to our problem as 
\begin{equation}
\delta E^{\rm F} =-\int \frac{dq}{2\pi}
\left(1- \frac{\varepsilon'\left(p\right)}{\varepsilon'\left(q^{*}\right)}\right)
e^{-2 i q^{*} L} S_{0}(q^{*},p) S_{0}(p,-q^{*}) f(q^{*},p)^2 \label{lus1}
\end{equation}
where $f(q^{*},p)$ is a function determined by the $S$-matrix elements between
the physical particle and virtual (mirror) particles:\footnote{Note that we considered both of right-moving and left-moving of virtual particles.} 
\begin{equation}
f(q^{*},p) = 2 a_1(q^{*},p) a_1(p, -q^{*}) + a_2(q^{*},p) a_2(p, -q^{*}) - 2 a_6(q^{*},p) a_5(p, -q^{*}) 
\end{equation}
Here, $q$ and $q^{*}$ are separately the energy and momenta of the virtual
particles and they satisfy the on-shell relation $q^2 + \epsilon^2(q^{*}) =
0$. 
The functions appearing in (\ref{lus1}) are given as follows \cite{Arutyunov:2006yd}:\begin{eqnarray}
S_{0}(p_1,p_2) &=& \frac{x_{p_2}^{+} - x_{p_1}^{-}}{x_{p_2}^{-} - x_{p_1}^{+}} \frac{1-\frac{1}{x_{p_2}^{-} x_{p_1}^{+}}}{1-\frac{1}{x_{p_2}^{+} x_{p_1}^{-}}} \sigma^2 (p_1,p_2) \cr
a_{1} (p_1,p_2)&=& \frac{x_{p_2}^{-} - x_{p_1}^{+}}{x_{p_2}^{+} - x_{p_1}^{-}} \sqrt{\frac{x_{p_2}^{+}}{x_{p_2}^{-}}} \sqrt{\frac{x_{p_1}^{-}}{x_{p_1}^{+}}} \cr
a_{2} (p_1,p_2)&=& \frac{\left(x_{p_1}^{-} - x_{p_1}^{+}\right)\left(x_{p_2}^{-} - x_{p_2}^{+}\right)
  \left(x_{p_2}^{-} - x_{p_1}^{+}\right)}{\left(x_{p_1}^{-} - x_{p_2}^{+}\right)
  \left(x_{p_2}^{-}x_{p_1}^{-} - x_{p_2}^{+}x_{p_1}^{+}\right)}
  \sqrt{\frac{x_{p_2}^{+}}{x_{p_2}^{-}}} \sqrt{\frac{x_{p_1}^{-}}{x_{p_1}^{+}}} \cr
a_{5} (p_1,p_2)&=& \frac{x_{p_1}^{-} - x_{p_2}^{-}}{x_{p_1}^{-} - x_{p_2}^{+}} \sqrt{\frac{x_{p_2}^{+}}{x_{p_2}^{-}}} \cr
a_{6} (p_1,p_2)&=& \frac{x_{p_1}^{+} - x_{p_2}^{+}}{x_{p_1}^{-} - x_{p_2}^{+}} \sqrt{\frac{x_{p_1}^{-}}{x_{p_1}^{+}}}
\end{eqnarray}
We note that the contribution from the reflection matrix $\mathbb{R}$ (including $\sigma(q^{*},-q^{*})$) is cancelled between left and right boundaries at the leading order. 
Then, one can straightforwardly check that (\ref{quaneff}) is equivalent to (\ref{lus1}) as we have $x_{q^{*}} \simeq x$ and $x_{p}^{\pm} \equiv X^{\pm}$ in the scaling limit\footnote{We used here $X^{\pm}=\frac{1}{X^{\mp}}$ because we consider the non-dyonic, simple boundary magnon.}.
Finally, one can express the one-loop energy shift from (\ref{bgm}) as:
\begin{equation}
\delta \epsilon_{\rm 1-loop} = \int \frac{dx}{\pi i} \frac{32 x^3}{(x^2 -1)^2}   \frac{(X^{+}-X^{-})^2 e^{\frac{-4 i x \Delta }{g(x^2 -1)}}}{(x-X^{+})(x+X^{-})(X^{-}x -1 ) (X^{+}x+1)} \label{oneloope}
\end{equation}
Here, we omitted the second term in bracket of (\ref{flucfre}) as it is suppressed  at the saddle point $x= i$.

\section{Conclusions}

In this paper we considered the generalization of the spectral curve from 
closed strings to open ones. We defined this spectral curve with the aid
of the logarithms of the eigenvalues of the open monodromy matrix - the
supertrace of which is the generator of conserved quantities. We showed that
this definition makes possible to determine all the analytic properties of the
spectral curve in the same way as in the case of closed strings and emphasized
the consequences of the additional (\lq\lq reflection'') symmetry that is
absent for closed strings.  

We analyzed this spectral curve from different points of view in case of open
strings attached to the $Y=0$ brane. First, from first principles, we
determined the explicit form of the spectral curve for some simple classical
open string solutions. Then, exploiting that for the $Y=0$ brane both the ABA
and the asymptotic $Y$ system solutions are available, we derived and
characterized the curve as the appropriate scaling limit of these
solutions.
 Finally we showed on two explicit examples 
how the spectrum of small fluctuations around a
classical solution can be determined by appropriately modifying the well known
procedure of the periodic case.

\begin{figure} 
\begin{center}
\subfigure[]{\includegraphics[angle=0,width=0.45\textwidth]{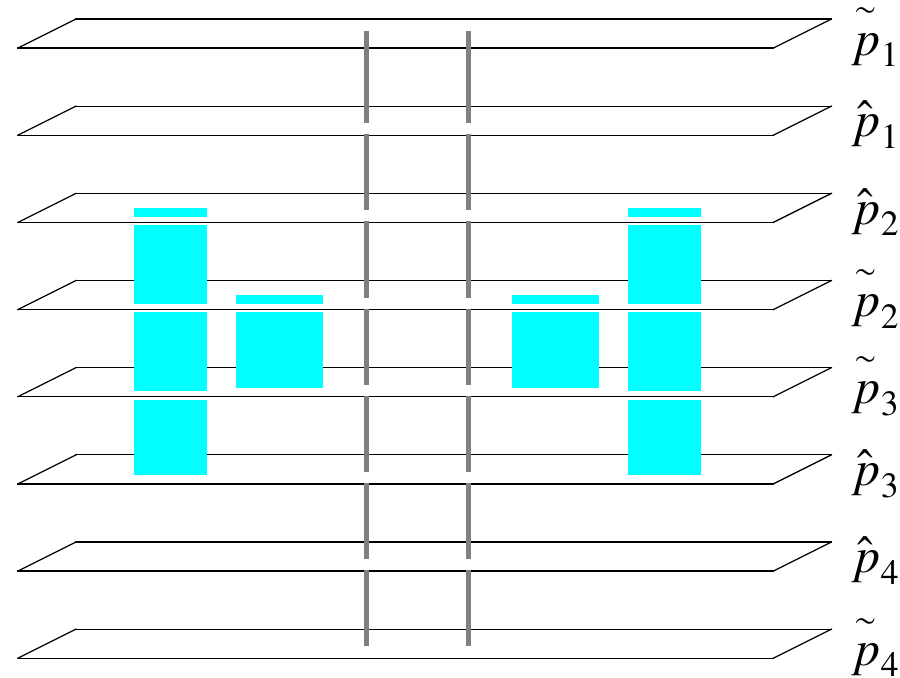}} 
\hspace{1cm} 
\subfigure[]{\includegraphics[angle=0,width=0.45\textwidth]{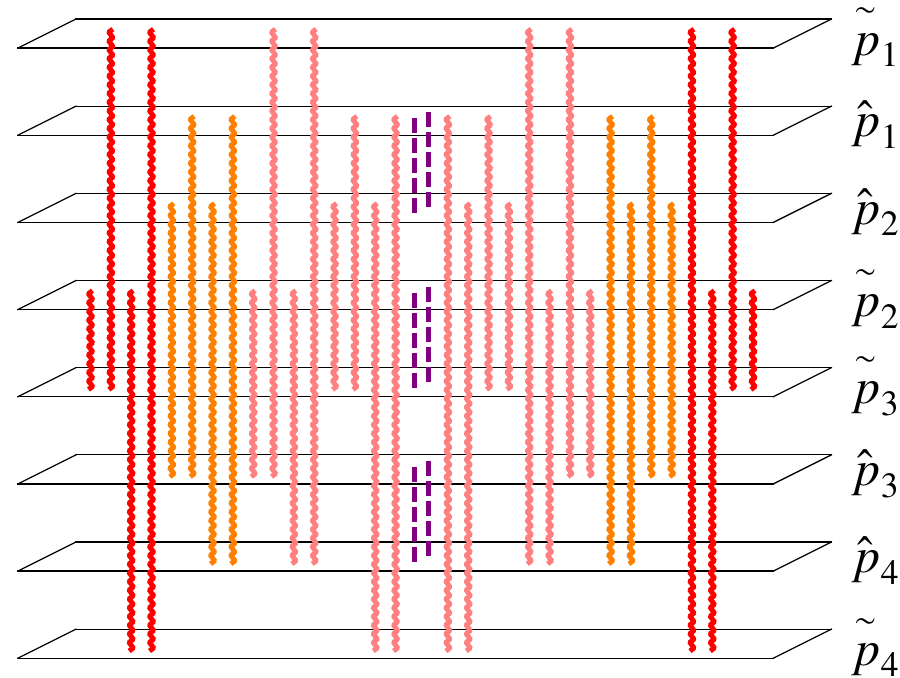}}
\caption{(a) General open string spectral curve is shown. It has poles at $x=\pm 1$ indicated by gray lines and cuts from various resolvents, which come in symmetric pairs due to the reflection symmetry. (b) Quasiclassical fluctuations for $Y=0 $ spectral curve are shown.  The boundary condition results in poles connecting three pairs of surfaces at $x=\pm i$ (purple dashed lines). Open string fluctuations are related to additional poles connecting each Riemann sheets at $x=\pm x_{n}^{i j}$ in a symmetric way. Wavy lines with red, orange and pink color represent 16 polarizations: four $S^5$ modes,  four $AdS_5$ and eight fermionic modes, respectively.   
\label{bdrycurve}}
\end{center}
\end{figure}

The consistent picture emerging from this series of investigations is that the
quasimomenta of the open case are very similar to that of the closed string
case. The differences - that mainly arise as a result of the reflection
symmetry - are that the residua of the poles at $x=\pm 1$ appearing in the
various quasimomenta are more tightly related to each other than in the
periodic case and that the  resolvent densities describing the various
excitations come in the form of symmetric pairs (e.g. $H_-(x)+H_+(x) $ with poles at $x =  x_{j}$ and  $x = - x_{j}$, respectively). Also
we found that the presence of the boundary gives a new pole, 
$B(x)=\frac{1}{2g}\frac{x}{x^2+1}$, to all
of the quasimomenta as a quantum effect. See figure \ref{bdrycurve}. 

This  boundary contribution is specific to the \(Y=0\) brane boundary conditions and shows up as a sub-leading quantum effect at the classical string regime while as a  finite-size effect at one-loop gauge theory regime \cite{Chen:2004mu, Chen:2004yf, Okamura:2005cj}, see Appendix \ref{sec:BA}.  By interpolating the quasimomenta from string theory to gauge theory,  we can see that the physical poles at the imaginary coordinates $x= \pm i$ in $B(x)$ are unified into a single pole at $x=0$ in gauge theory.   
Even though we ignored such a boundary contribution when we computed the semiclassical correction to the  BMN state and to the  boundary giant magnon, it was consistent  with  the L\"uscher's leading $F$-term result. However, the boundary contribution is expected to show up as a subleading quantum correction. 
It would be important to confirm this by computing subleading corrections. 

In this paper we derived the general properties of the boundary spectral curve and investigated it for  open strings satisfying the \(Y=0\) brane boundary conditions. But ABA  equations are  also known for many other integrable boundary conditions \cite{Correa:2013em,Correa:2012hh,Correa:2011nz,Correa:2009dm,
Drukker:2012de} and it would be interesting to extend our analysis for those cases. Especially to calculate the spectral curve for the \(q\bar q\) potential or for the vacuum expectation values of Wilson loops from first principles, since in  \cite{Janik:2012ws,Dekel:2013dy,Dekel:2013kwa} the  authors used the Lax matrix instead of the monodromy matrix, while in \cite{Gromov:2012eu, Gromov:2013qga, Sizov:2013joa} they analyzed the classical limit of the near BPS FiNLIE formulations to define such a curve.   

Recently the quantum spectral curve was proposed in the periodic $AdS_5
/CFT_4$ context \cite{Gromov:2013pga}. As such a quantum curve has  the entire information for  the full quantum spectrum, constructing the quantum curve for the  boundary problem would be an interesting direction for future research. 

\section*{Acknowledgements}

We would like to thank Piotr Surowka for his collaboration on the initial stage of this project and useful discussions.
Z.B. thanks Amit Dekel, Romuald Janik and Volodya Kazakov for enlightening discussions and
Romuald Janik for sharing his notes on the explicit calculations of the quasimomenta for periodic circular strings. Z.B. and L.P. were supported in part by 
OTKA 81461 and Z.B. also by a Lend\"ulet Grant. M.K. was partly supported by the National Research Foundation of Korea (NRF) grant funded by the Korea government (MEST)
through the Center for Quantum Spacetime (CQUeST) of Sogang University with grant number 2005-0049409,  Hungarian scholarship (type D) through the Balassi Institute and WCU Grant No. R32-2008-000-101300.

\appendix

\section{Transportation matrix  and  $U$ for  the $Y=0$ brane}\label{sec:Uexplicit}
We pointed out that $T(\zeta)$ can be defined from the transport matrix and the $U$ matrices. 
In \cite{Dekel:2011ja} the $U$ matrix corresponding to the $Z=0$ case
was determined while here we need it for the $Y=0$ one. We construct it 
below by solving all the necessary conditions listed in \cite{Dekel:2011ja} rather then
trying to rotate the result of \cite{Dekel:2011ja} in an appropriate way.

To get the desired metric we use the coset element representative
$g=g_{AdS_5}g_{S^5}$ with
\begin{equation}
g_{AdS_5}=e^{P_0t}e^{-J_{13}\phi}e^{J_{24}\Phi}e^{-J_{14}\alpha}e^{P_1\rho},\quad\quad 
g_{S^5}=e^{-J_{79}\phi_1}e^{P_8\phi_2}e^{J_{56}\phi_3}e^{P_6\psi}e^{P_7(\pi/2-\gamma)}.
\end{equation}
The giant graviton corresponding to the $Y=0$ brane is given by Dirichlet boundary conditions $\psi=0$, $\rho=0$ together with Neumann
boundary conditions for the rest of the coordinates 
$\partial_\sigma\gamma=\partial_\sigma \phi_2=\partial_\sigma\phi_1=0$. At the
boundary, the bosonic sectors current, $A^{(2)}$, has the following world sheet
components
\begin{equation}
A^{(2)}_\tau=P_0\partial_\tau
t-P_7\partial_\tau\gamma+P_8\sin\gamma\partial_\tau\phi_2+P_9\cos\gamma\partial_\tau\phi_1,\quad  
A^{(2)}_\sigma=P_1\partial_\sigma\rho+P_6\sin\gamma\partial_\sigma\psi.
\end{equation}
Therefore the natural Ansatz for the $U$ matrix is $U=aP_0+bP_7P_8P_9$ with
constant $a,\ b$ to be determined and plugging this into the conditions listed
in \cite{Dekel:2011ja} we found that up to normalization and relative sign 
\begin{equation}
U=2P_0-i2^3P_7P_8P_9.
\label{ourU}
\end{equation}

In the paper we used the following conventions of $P_i$ matrices.
The $so(4,1)$ generators $P_0,\dots ,P_4$ are described as
\begin{equation}
P_0,\dots ,P_4=\begin{pmatrix}
0_{4\times 4} & \ \\ 
\  & \frac{i}{2}\gamma^5,\frac{1}{2}\gamma^1,\dots ,\frac{1}{2}\gamma^4
\end{pmatrix}
\end{equation}
while the $so(5)$ generators $P_5,\dots ,P_9$ as 
\begin{equation}
P_5,\dots ,P_9=\begin{pmatrix}
\frac{i}{2}\gamma^1,\dots ,\frac{i}{2}\gamma^5 & \ \\ 
\  & 0_{4\times 4}
\end{pmatrix}
\end{equation} 
in terms of the $4\times 4$ 
Dirac matrices $\gamma^i$ ($i=1,\dots 5$):  $\gamma^5={\rm diag}(1,1,-1,-1)$
\begin{equation}
\gamma^1=\begin{pmatrix} 0 & 0& 0 &-1\\
                         0 & 0& 1 & 0\\
                         0 & 1& 0 & 0\\
                         -1& 0& 0 & 0\\
\end{pmatrix} \quad
\gamma^2=\begin{pmatrix} 0 & 0& 0 &i \\
                         0 & 0& i & 0\\
                         0 & -i& 0 & 0\\
                         -i& 0& 0 & 0\\
\end{pmatrix} \quad    
\gamma^3=\begin{pmatrix} 0 & 0& 1 & 0\\
                         0 & 0& 0 & 1\\
                         1 & 0& 0 & 0\\
                         0 & 1& 0 & 0\\
\end{pmatrix} \quad
\gamma^4=\begin{pmatrix} 0 & 0& -i& 0\\
                         0 & 0& 0 & i\\
                         i & 0& 0 & 0\\
                         0 &-i& 0 & 0\\
\end{pmatrix} 
\end{equation}
satisfying the Clifford algebra $\{\gamma^i,\gamma^j \}=2\delta^{ij}$. Using
these explicit expressions in (\ref{ourU}) gives 
\begin{equation}
U=i\ {\rm diag}(1,-1,1,-1,1,1,-1,-1)\;.
\label{Uexplicit}
\end{equation}

\section{Eigenvalues of       $T_{}(x)$ for circular strings with $n=2N$}\label{sec:Teigenv}
In this appendix we solve the differential equation
$\partial_\sigma\psi=H\psi$ and determine the $S^5$ eigenvalues of
$T(x)$. First,  
by a constant similarity transformation - we bring $H$ in (\ref{S5corner}) to the form
$H\rightarrow \tilde{H}=\begin{pmatrix}i\tilde{b} & 0\\
                                        0 & -i\tilde{b}\end{pmatrix}$ and
solve 
\begin{equation}
\partial_\sigma\psi=\pm i\tilde{b}\psi, \qquad \psi =\begin{pmatrix}\psi_1\\
                                                                   \psi_2\end{pmatrix}.\label{diffegy}\end{equation}
Taking the upper sign we recognize that the equations can be solved by making
the Ansatz\footnote{We thank Romuald Janik for sharing his explicit calculation for the quasimomenta of the analogous circular string solution in the periodic case. }
\begin{equation}
\psi_1-\psi_2=Ae^{\alpha\sigma},\qquad\quad \psi_1+\psi_2=Be^{\gamma\sigma},
\end{equation}
if $\gamma -in=\alpha $ holds. Furthermore using this in the quadratic
equation guaranteeing that we may have non trivial $A$ and $B$ determines
$\gamma $ as
\begin{equation}
\gamma_{1,2}=\frac{i}{2}n\pm i\frac{x}{x^2 -1}\sqrt{\frac{n^2}{x^2}+w^2}.
\end{equation}
Repeating this procedure with the lower sign in eq.(\ref{diffegy}) gives
\begin{equation}
\gamma_{3,4}=\frac{i}{2}n\pm i\frac{x}{x^2 -1}\sqrt{n^2x^2+w^2}
\end{equation}
As the eigenvalues of $\tilde{t}(\pi ,0,x)$ can be obtained as the ratios 
$\psi(\pi )_j/\psi(0)_j$ 
in both cases we compute  
\begin{equation}
\frac{\psi_1(\pi
  )}{\psi_1(0)}=e^{\gamma\pi}\frac{Ae^{-in\pi}+B}{A+B}=e^{\gamma\pi},\qquad
 \frac{\psi_2(\pi
   )}{\psi_2(0)}=e^{\gamma\pi}\frac{B-Ae^{-in\pi}}{B-A}
 =e^{\gamma\pi},
\end{equation}
(In the last equality we exploited that $n=2N$). Since
$\tilde{t}(\pi,0,-x)^{-1}$ has the same diagonal form as $\tilde{t}(\pi,0,x)$
eventually we find that $T(x)$'s eigenvalues in the $S^5$ corner are
$-e^{2\pi\gamma_{1,2}}$ and $-e^{2\pi\gamma_{3,4}}$.

\section{Bethe roots, Dynkin labels and conserved charges}\label{sec:Dynkin}
In this Appendix we analyze the asymptotics of the quasimomenta. By keeping the leading order term of the quasi-momenta, (\ref{bdryqm}),  in the large $x$ limit we find \begin{eqnarray}
{\hat p}_1 &=& \frac{J+Q_2 -K_1 +2 K_2 -K_3 +K_4}{g x}, \,\ \quad {\tilde p}_1 = \frac{J-K_1-K_3+K_4}{g x} \cr
{\hat p}_2 &=& \frac{J+Q_2 +K_1 -2 K_2 +K_3 +K_4}{g x}, \,\ \quad {\tilde p}_2 = \frac{J+K_1+K_3-K_4}{g x} \cr
{\hat p}_3 &=& -\frac{J+Q_2 +K_4 +K_5 -2K_6 +K_7}{g x},  \quad {\tilde p}_3 = -\frac{J-K_4+K_5+K_7}{g x} \cr
{\hat p}_4 &=& -\frac{J+Q_2 +K_4 -K_5 +2K_6 -K_7}{g x},  \quad {\tilde p}_4 = -\frac{J-K_4+K_5+K_7}{g x}, \label{asymp}
\end{eqnarray} 
where $Q_2 = \delta \Delta$ is the second conserved charge - energy.
We choose the gradings $\eta_1 = \eta_2 =1$ and the unphysical hypercharge $B=0$, such that the numbers of the  Bethe roots $K_{j}$ can be expressed in terms of the Dynkin labels of $SU(2,2)$ and $SU(4)$ ,  given by $[q_1,p,q_2]$ and $[s_1,r,s_2]$ as follows \cite{Beisert:2005fw}:
\begin{eqnarray}
K_1 &=& \frac{1}{2} J - \frac{1}{4}(2 p+3 q_1 +q_2) \cr 
K_2 &=& -\frac{1}{4} \left(2(r+Q_2) +3 s_1 +s_2 +2 p +3 q_1 +q_2 \right) \cr
K_3 &=& -\frac{1}{2} J - \frac{1}{2} \left(2(r+Q_2) - s_1 +s_2\right) -s_1   -\frac{1}{4}(2 p- q_1 +q_2) -q_1 \cr
K_4 &=& -r-Q_2-\frac{1}{2}(s_1 +s_2 + q_1 +q_2 )-p \cr
K_5 &=& -\frac{1}{2} J - \frac{1}{2} \left(2(r+Q_2) + s_1 -s_2\right) -s_2   -\frac{1}{4}(2 p+ q_1 -q_2) -q_2 \cr
K_6 &=& -\frac{1}{4} \left(2(r+Q_2) +s_1 +3 s_2 +2 p + q_1 +3 q_2 \right)\cr
K_7 &=& \frac{1}{2} J - \frac{1}{4}(2 p+ q_1 +3 q_2) \label{bethedynkin} 
\end{eqnarray}
Since each Dynkin labels are related to the conserved charges as \cite{Beisert:2005bm}  
\begin{eqnarray}
&& q_1 = J_2 - J_3,\quad p = J_1 - J_2,\quad q_2 = J_2 + J_3, \cr
&& s_1 = S_1 - S_2,\quad r = -\Delta - S_1,\quad s_2 = S_1 + S_2, \label{dynkincharge}
\end{eqnarray}
we finally obtain the large $x$ asymptotics (\ref{largexasymp}) of our quasi-momenta.

\section{Spectral curve from the one loop Bethe ansatz}\label{sec:BA} 

The asymptotic limit of the one-loop BA equation for the open case was analyzed in \cite{Chen:2004mu, Chen:2004yf}, while the spectral curve was
proposed in this context in \cite{Okamura:2005cj}. For 
completeness we summarize their findings.

 The $Y$ function in the asymptotic limit simplifies to 
\begin{equation}
Y_{1,0}=\left(\frac{u+i/2}{u-i/2}\right)^{2L} \frac{u-i/2}{u+i/2}\prod_{j=1}^{N}\frac{u-u_j+i}{u-u_j-i}
\frac{u+u_j+i}{u+u_j-i} . 
\end{equation}
This $Y$ function is the same as the periodic one with particle content $(u_j,-u_j)$ in volume $2L$, 
with the exception of the factor $u^-/u^+$. This factor, when evaluated at $u_j$ in the BA equation
 (\ref{YBA}),  is responsible for removing the unwanted selfscattering piece. This will not be 
relevant in the scaling limit, in which  $L\to\infty$ and roots scale as $u_{j}\propto L$. 
In this limit we reparametrize
them as $u_{j}=L x_{j}$, and expand the logarithm of the BA equations
for large $L$: 
\begin{equation}
\frac{1}{x_{j}}=\frac{1}{L}\sum_{k:k\neq j}^{N}\left(\frac{1}{x_{j}-x_{k}}+\frac{1}{x_{j}+x_{k}}\right)-2\pi n_{j}
\end{equation}
Clearly if $x_{j}$ is the solution of the equation with $n_{j}$
then $-x_{j}$ is a solution with $-n_{j}$. In the $L\to\infty$
limit roots condense on symmetric cuts localized around $\pm 1/(2\pi n_{j})$. We
introduce their densities and resolvents as
\begin{equation}
\rho(x)=\frac{1}{L}\sum_{k}^{N}\delta(x-x_{j})
\quad ; \quad 
G(x)=\frac{1}{L}\sum_{k}^{N}(\frac{1}{x-x_{j}}+\frac{1}{x+x_{j}})=\int_{C}dx'\frac{\rho(x')}{x-x'}
\end{equation}
which are nonzero on cuts $C_{\alpha}^{\pm}=\pm( a_{\alpha}, b_{\alpha})$
with $a>0$ and $b>0$.
The resolvent is an analytic function on the complex plane with given cuts and has
the asymptotics ($x\to\infty$):
\begin{equation}
G(x)=\frac{2\alpha}{x}+\dots\qquad;\qquad\int_{C}dx\rho(x)=2\alpha=\frac{2N}{L}
\end{equation}
The  quasi momenta are related to the resolvent in a trivial way
\begin{equation}
p(x)=G(x)-\frac{1}{2x}
\end{equation}
such that the BA equation takes the form
\begin{equation}
p(x+i0)+p(x-i0)=\pm2\pi n_{k}\quad;\qquad x\in C_{k}^{\pm}
\end{equation}
Similarly to the periodic case $p(x)=\int^{x}dp$ is an Abelian integral
for the meromorphic differential $dp$, which has two double poles
at $x=0$ and integer periods
\begin{eqnarray}
2\pi(n_{k}-n_{j}) & = & p(x_{k}+i0)-p(x_{j}-i0)+p(x_{k}-i0)-p(x_{j}+i0)=\nonumber \\
 & = & \int_{x_{j}-i0}^{x_{k}+i0}dp+\int_{x_{j}+i0}^{x_{k}-i0}dp=\oint_{B_{ij}}dp
\end{eqnarray} on a hyperelliptic curve: 
\begin{equation}
y^{2}=\prod_{k}^{2N}(x-x_{k})(x+x_{k})\qquad C_{k}^{+}=\{x_{2k},x_{2k+1}^{*}\}
\end{equation}

Comparing these results to the periodic case we observe that the only difference is
that cuts appear in the classical limit in a symmetric way.

\bibliographystyle{JHEP}

\end{document}